\documentclass[12pt,a4paper,fleqn]{article}

\usepackage[textheight=23cm,textwidth=15cm]{geometry}
\usepackage{amsmath}
\usepackage[version=3]{mhchem}
\usepackage{graphicx}
\usepackage[american]{babel}
\usepackage{here}
\usepackage[articletitle=true, maxauthors=150]{achemso}
\usepackage{siunitx}
\usepackage{multirow}

\usepackage{url}
\usepackage{xcolor}

\usepackage[affil-it]{authblk}

\usepackage{xspace}
\begin{document}

\begin{center}

{\LARGE\bf
Automated Preparation of Nanoscopic Structures: Graph-Based Sequence Analysis, Mismatch Detection, and pH-Consistent Protonation with Uncertainty Estimates \\}

\vspace{1cm}

\renewcommand*{\thefootnote}{\fnsymbol{footnote}}

{\large
Katja-Sophia Csizi
and
Markus Reiher\footnote{Corresponding author; e-mail: mreiher@ethz.ch}
}\\[4ex]

\renewcommand*{\thefootnote}{\arabic{footnote}}
\setcounter{footnote}{0}

ETH Zurich, Department of Chemistry and Applied Biosciences, Vladimir-Prelog-Weg 2,
8093 Zurich, Switzerland 

\vspace{.5cm}
November 10, 2023

\vspace{.43cm}

\textbf{Abstract} 
\end{center}
\vspace{-0.5cm}
  
\vspace*{-.41cm}
{\small
Structure and function in nanoscale atomistic assemblies are tightly coupled, and every atom with its specific position and even every electron will have a decisive effect on the electronic structure, and hence, on the molecular properties.
Molecular simulations of nanoscopic atomistic structures therefore require accurately resolved three-dimensional input structures. If extracted from experiment, these structures often suffer from severe uncertainties, of which the lack of information on hydrogen atoms is a prominent example. Hence, experimental structures  require careful review and curation, which is a time-consuming and error-prone process. 
Here, we present a fast and robust protocol for the automated structure analysis and  pH-consistent protonation, in short, ASAP. For biomolecules as a target, the ASAP protocol integrates sequence analysis and error assessment of a given input structure. ASAP allows for  p$K_\mathrm{a}$\xspace prediction from reference data through Gaussian process regression including uncertainty estimation and connects to system-focused atomistic modeling described in (J. Chem. Theory Comput. 16, 2020, 1646). Although focused on biomolecules, ASAP can be extended to other nanoscopic objects, because most of its design elements rely on a general graph-based 
foundation guaranteeing transferability. The modular character of the underlying pipeline supports different degrees of automation, which allows for (i) efficient feedback loops for human-machine interaction with a low entrance barrier and for (ii) integration into autonomous procedures such as automated force field parametrizations. This facilitates fast switching of the pH-state through on-the-fly system-focused reparametrization during a molecular simulation at virtually no extra computational cost. 
}
\newpage

\section{Introduction}
\label{sec:introduction}
The generation of a high-fidelity initial 
structure of a nanoscopic chemical system for atomistic modeling 
is an important, but nontrivial task. Especially for functional biopolymers, for which extensive experimental structural databases exist, minimal structural rearrangements within an active site or through allosteric effects can change, drive, or even impede chemical processes such as enzyme catalysis.\cite{Wodak2019, Zhou2019}
 
Of special usefulness and need is a protein-oriented framework that enables control and building of structural models in a fast, reproducible, robust, physically realistic, standardized, and automated manner, making it possible to process structures with minimal and unbiased human intervention. Only by such frameworks, computational studies of large structures (for instance, relying on molecular quantum mechanics (QM), molecular mechanics (MM), QM/MM hybrid simulations, or molecular dynamics (MD)) can be efficiently carried out in a comprehensive and reproducible way.\cite{Garcia-Viloca2004} 
While efforts have been made to predict structures from sequence information alone~\cite{Senior2020, Jumper2021, Baek2021}, most simulations rely on experimentally curated input structures taken from X-ray crystallography or NMR spectroscopy, which are collected in the protein data bank (PDB)~\cite{Berman2000}. 

Leveraging experimental data for simulations is appealing and provides dense molecular information such as coordinates, atom types, residue and chain information, connectivity, conformers, and more. However, these structures often suffer from significant uncertainties in the atomic coordinates and other structural errors. For instance, incompletely resolved residues, unreasonable connectivities, and atomic clashes can occur. 
These mismatches must be detected and corrected, which can be handled in an automated fashion for most of the above-mentioned structural issues through implementation of simple chemical rules. Chemical valences of an atom may, for example, be evaluated by deriving connectivity information from simple distance-based rules, at least for organic and main-group chemical compounds, which make up for a large part of many nanoscopic bio-macromolecules.

Even accurately resolved experimental data that are not plagued by these mismatches require careful processing. 
For instance, multiple conformers or ensembles of molecular models (determined by NMR experiments)
of the same structure can be condensed into a single PDB file depending on the time resolution of the underlying measurements. For enzymes, the conformers that reflect chemically relevant structural intermediates of a catalytic cycle might be only a small portion of the recorded structures in the file. 
Moreover, proteins may be subjected to modifications (such as mutations) to facilitate the crystallization process,\cite{Longenecker2001} which might require correction for simulations of generic parent systems.

The multitude of the above-mentioned structural uncertainties and mismatches are challenging to detect but propagate into the property prediction in a simulation, because every single atom and its position will have a decisive effect on the electronic structure of the compound. Therefore, it is vital to review carefully every molecular structure prior to simulation. 

Apart from automated mismatch detection, the second challenge of structure evaluation is that X-ray diffraction usually lacks resolution of hydrogen atoms (which often can only be recovered by more advanced techniques such as neutron scattering), and the many vacant valences need to be occupied in a 
chemically consistent way, reflecting also chemical constraints such as pH value and solvation. 
Whereas the overall pH-stability of a nanoscopic system is primarily determined by the global p$K_\mathrm{a}$\xspace value of the macromolecule, specific residues adopt site-specific p$K_\mathrm{a}$\xspace values based on their local environment, guiding the protonation state of this site.
However, the assignment of meaningful discrete protonation configurations in a protein is by no means trivial, because the many close and long-range electrostatic interactions that affect the protonation state of one specific site must be accurately captured by the model that is employed to predict the negative logarithm of the acid dissociation constant, the p$K_\mathrm{a}$\xspace. Although the protonated configuration of highest probability could, in principle, be found by extensive search in the space of protonated configurations, this will rapidly become unfeasible because of the combinatorial growth of their number with the number of protonation sites and the number of protons to be added.

In this work, we present a physically meaningful and robust pipeline for \textit{a}utomated \textit{s}tructure \textit{a}nalysis and pH-consistent \textit{p}rotonation (denoted \textit{ASAP}) based on quantum chemical reference data. 
Our pipeline combines sequence analysis of a protein, mismatch detection, addition of hydrogen atoms, solvation, and p$K_\mathrm{a}$\xspace prediction from a quantum reference. It has been implemented as part of our open-source software project \textsc{Swoose}~\cite{swoose1.0.0} within the \textsc{Scine}~\cite{scine} infrastructure. While there are multiple computer programs that can complete individual steps of the structure preparation pipeline, only few unify many of the ASAP features, for instance, the protein preparation wizard\cite{Sastry2013} and the \textsc{PdbFixer} of OpenMM\cite{openMM}.
Compared to them, our ASAP pipeline offers additional features as described in the following.

We accomplish error detection through a graph-isomorphism ansatz that extracts faulty occurrences of structural building blocks within a large structure. We derive protonation probability through a machine learning approach. In particular, a regression ansatz based on a Gaussian process is trained on experimental p$K_\mathrm{a}$\xspace values and corresponding quantum chemical reference data. As any physicochemical property model should be equipped with confidence intervals~\cite{Proppe2016, Simm2017}, we provide uncertainty estimates for predicted p$K_\mathrm{a}$\xspace values directly delivered by the Gaussian process, such that special consideration is given to residues whose p$K_\mathrm{a}$\xspace values are close to the system pH because they are more prone to be misclassified. This delivers p$K_\mathrm{a}$\xspace values from the deprotonation free energy in solution, which can be obtained with a quantum chemical method.

Our workflow is seamlessly integrated into our pipeline for automated construction of quantum classical hybrid models.\cite{Brunken2021} This implies that the system-focused atomistic model~\cite{Brunken2020} (SFAM) for the given structure can be generated, and the corresponding quantum region can be selected in an automated fashion.\cite{Brunken2021} 
The design of ASAP also enables automated, on-the-fly (re)-parametrization of the SFAM force field for a specific protonation state configuration, which we will discuss in detail in future work. 
Although our approach is optimized for biomolecular systems, it can be extended to other nanoscopic architectures due to its graph-based and first-principles foundations, both being agnostic to the specific 
atomic composition. 

This work is organized as follows: In section \ref{sec:theory}, we discuss how machine learning protonation probabilities can be derived from quantum chemical reference data and corresponding confidence intervals from Gaussian processes, followed by an overview of the computational methodology applied. In section \ref{sec:asap}, we present the conceptual design and all components of the ASAP pipeline.
In section \ref{sec:results}, we demonstrate the structure preparation protocol for chain A of the peptide hormone insulin and subsequently study our approach for a set of protein structures extracted from the PDB. Then, we illustrate that p$K_\mathrm{a}$\xspace values in solution can be derived from a 
linear regression ansatz, for which the isolated amino acids represent an ideal test case as their p$K_\mathrm{a}$\xspace values are well known from experiment. Finally, we derive protonation probabilities for selected sites in chain A of insulin.

\section{General Considerations and Computational Methodology}
\label{sec:theory}
\subsection{Protonation Probabilities from a Quantum Reference}
\label{subsec:pka_from_quantum_reference}
The p$K_\mathrm{a}$\xspace value is a quantitative measure of the  propensity of an acidic molecule AH to dissociate into a proton ($\mathrm{H^{+}}$) and the corresponding base ($\mathrm{A^{-}}$) in aqueous solution, measuring the protonation probability at a given pH. It is defined as the negative base-10 logarithm of the acidity constant $K_a$, which is the ratio of the concentrations of proton H$^+$ and base A$^-$ and the acid AH in solution:
\begin{equation}
    pK_a = -\log_{10} (K_a) = -\log_{10}\frac{[A^-][H^+]}{[AH]}.
    \label{eq:equilibrium}
\end{equation}
Note that Equation (\ref{eq:equilibrium}) is only valid at low proton concentrations. In cases of high proton concentrations, the p$K_\mathrm{a}$\xspace value should be expressed as a function of activities and activity coefficients of the respective species instead of concentrations.
The acidity constant relates to the Gibbs free energy $\mathrm{\Delta G_a}$ of AH deprotonation in solution:
\begin{equation}
    K_a = \exp\Big( - \frac{\Delta G_a}{RT}\Big). 
    \label{eq:ka-gibbs}
\end{equation}
${\Delta G_a}$ in solution is given by\cite{Alongi2010}:
\begin{equation}
 {\Delta G_a = G^{\mathrm{aq}}(A^-) + G^{\mathrm{aq}}(H^+) - G^{\mathrm{aq}}(AH) = \Delta G^{\mathrm{aq}}_0 + G^{\mathrm{aq}}(H^+)} 
 \label{eq:Ga}
\end{equation}
with $G^{\mathrm{aq}}(H^+)$, $G^{\mathrm{aq}}(A^-)$, and $G^{\mathrm{aq}}(AH)$ being the free energy of a proton, a base $A^-$, and the corresponding acid $(AH)$ in aqueous solution. The difference of the latter two is denoted $\Delta G^{\mathrm{aq}}_0$.
Insertion of Equation (\ref{eq:ka-gibbs}) into Equation (\ref{eq:equilibrium})
yields\cite{Alongi2010}
\begin{equation}
    {pK_a = \frac{\Delta G_a}{R T \ln(10)}}.
    \label{eq:pka}
\end{equation}
For a reliable prediction of p$K_\mathrm{a}$\xspace values, the energy contributions in Equation (\ref{eq:Ga}) must be calculated with sufficient accuracy, because an error of $\mathrm{\sim}$ 1.36 kcal $\mathrm{mol^{-1}}$ in ${\Delta G_a}$
translates into one p$K_\mathrm{a}$\xspace unit of error on the logarithmic scale at $T=$293.15 K.
This accuracy is hard to achieve
for standard quantum chemical methods such as density functional theory (DFT). Moreover, it is difficult to model the free energy of a proton in solution reliably (last term in Equation (\ref{eq:Ga})). Most studies rely on the indirect calculation of ${\Delta G_a}$ from a thermodynamic cycle 
(see, for instance, Refs. \citenum{Ho2009, Casasnovas2014, Ho2015, Seybold2015}). Alternatively, this quantity can be measured in experiments, which, however, can result in a mismatch with the theoretical model chosen for the calculation of the other free energy contributions.

Methods for p$K_\mathrm{a}$\xspace prediction can roughly be grouped into
microscopic, macroscopic, and empirical techniques. 
Macroscopic techniques are rooted in grid-based
continuum electrostatic schemes~\cite{Bashford1990, Wang2015a} solving the Poisson--Boltzmann equation. Limitations of the Poisson--Boltzmann ansatz are the neglect of specific protein-ion interactions and that properties are usually calculated for a single structure, which will only be valid if this structure is representative of protonated and deprotonated states.
Moreover, solvation effects described by a dielectric continuum in an implicit solvation model will lack the explicit description of hydrogen bonding between solute and solvent, although they can be crucial.\cite{Ho2014} 
Microscopic methods, by contrast, derive the p$K_\mathrm{a}$\xspace value from quantum chemical calculations,\cite{Klicic2002,Klamt2003, Riojas2014, Ho2015, Thapa2017, Lian2018} all-atom molecular dynamics simulations,\cite{Donnini2011, Itoh2011, Wallace2011, Huang2016a} or quantum classical hybrid models\cite{Li2002, Li2003}. For these methods, the different routes to p$K_\mathrm{a}$\xspace prediction 
build on the calculation of acidity in the gas or solution phase, either with a dielectric continuum or with an explicit solvent model.\cite{Ho2011} Empirical p$K_\mathrm{a}$\xspace prediction is mostly based 
on adding an empirical correction to a baseline model-p$K_\mathrm{a}$\xspace value~\cite{Godoy-Ruiz2005, Li2005, Krieger2006, Bas2008b}. This approach is valuable because of its unmatched readiness, allowing for p$K_\mathrm{a}$\xspace prediction in nanoscale structures within seconds, but it lacks a rigorous physical foundation. Various pipelines implement different routes towards p$K_\mathrm{a}$\xspace prediction (see, for example, \textsc{Pypka}~\cite{Reis2020},\textsc{PDB2PQR},\cite{Dolinsky2004, Dolinsky2007} and \textsc{DelPhi}~\cite{Wang2015a} for Poisson--Boltzmann-based 
p$K_\mathrm{a}$\xspace prediction, the empirical approaches 
${H++}$~\cite{Anandakrishnan2012} and \textsc{Propka},\cite{Olsson2011}, and \textsc{Jaguar}~\cite{Bochevarov2016a}, which is DFT-based). 

Klicic et al. have shown that errors in the computation of terms required in Equation (\ref{eq:pka}) are systematic~\cite{Klicic2002} (see also Refs. \cite{Alongi2010, Ho2014, Ratkova2015}), which can be exploited to leverage p$K_\mathrm{a}$\xspace prediction through a linear regression ansatz. In such an ansatz, the p$K_\mathrm{a}$\xspace value of a given site can be derived from free energies of acid and base only (condensed in ${\Delta G_0^\mathrm{aq}}$ on the right-hand side of Equation (\ref{eq:Ga})). These quantities can be calculated with standard electronic structure methods, whereas the fitting parameters absorb both, the value of $\mathrm{G(H^+)}$ and (ideally) all other deficiencies of the QM method employed: 
\begin{equation}
\begin{aligned}
    {pK_a} & = {\frac{s \Delta G_0^\mathrm{aq}}{ RT \ln(10)} + \frac{s G^\mathrm{aq}(H^+)}{RT \ln(10)}} 
                   =  {k\Delta G_0^\mathrm{aq} + C }.
    \end{aligned}
\label{eq:pka_from_linear_relationship}
\end{equation}
The scaling factor $s$ corresponds to an error correction for some given computational method.\cite{Matsui2012} 
An appropriate training set is then needed to derive the regression parameters $k$ and $C$. 
Not only does the approach then circumvent the need to calculate all quantities in the thermodynamic cycle, but it also delivers an estimate for the free energy of a proton for the applied 
method: 
\begin{equation}
    G^\mathrm{aq}(H^+) = \frac{C}{k}
\end{equation}

We apply Gaussian process (GP) regression to the regression problem in 
Equation (\ref{eq:pka_from_linear_relationship}) to derive confidence intervals for a given prediction. The data points comprise the ${\Delta G_0^\mathrm{aq}}$ values and the corresponding experimental p$K_\mathrm{a}$\xspace values. 
GP regression is a probabilistic Bayesian machine learning technique that  predicts continuous quantities from a given prior data set. An important advantage of GP models is that they deliver an estimate for the prediction uncertainty\cite{Pernot2015, Proppe2016, Proppe2017, Simm2016, Simm2017, Proppe2019, Reiher2021}, 
which can be leveraged for physicochemical models provided that suitable measures for molecular similarity can be found (cf., Refs. 
\cite{Simm2018, Proppe2019a, Deringer2021a, Gugler2022}).
For a theoretical introduction to GPs, we refer to Section SI in the Supporting Information. 
Although a linear relationship is known for our case, the application of GPs instead of ordinary linear regression provides several advantages here. First of all, the homoscedasticity assumption that the residual terms in ordinary linear regression have the same standard deviation is insufficient for the present problem. If our data point requires extrapolation from the range of training data, the uncertainty of the predictive mean will be high, resulting in an overall heteroscedastic error distribution. Second, GPs offer an easy way to learn (or incorporate) the uncertainty in both our variables \textbf{x} (precisely, the accuracy of the employed QM method to calculate ${\Delta G_0^\mathrm{aq}}$) and our function values \textbf{y} (the accuracy of the experimentally measured p$K_\mathrm{a}$\xspace value). The latter is defined by the hyperparameter ${\sigma_y^2}$ (for details, see Section SI in the Supporting Information). 
Note that Bayesian linear regression and bootstrapped linear least-squares regression~\cite{Proppe2017} are alternatives for the present regression problem.

\subsection{Computational Methodology}
For all DFT calculations with the \textsc{Orca} program package, version 4.2.0.\cite{Neese2012, Neese2020}
we applied the Perdew--Becke--Ernzerhof (PBE) pure~\cite{Perdew1996a} and the corresponding hybrid PBE0 hybrid~\cite{Ernzerhof1999, Adamo1999} density functionals with an Ahlrich's def2-TZVP basis set.\cite{Schafer1994} The resolution-of-the-identity density-fitting technique was selected to accelerate all PBE calculations.\cite{Eichkorn1995, Eichkorn1997}
Grimme's D3 dispersion correction with Becke--Johnson damping was turned on in all DFT calculations.\cite{Grimme2010a, Grimme2011} 
We applied the implicit solvation model SMD~\cite{Marenich2009} with water as solvent. The dielectric constant was chosen to be 80.4 and the solvent probe radius was set to 1.3 \r{A}. Thermodynamic quantities at 298.15 K were obtained by application of the standard gas-phase model of statistical 
thermodynamics;
that is, by approximating the translational, rotational, and vibrational degrees of freedom by the ideal gas, rigid rotor, and harmonic oscillator models, respectively. 
We applied our \textsc{Molassembler} library~\cite{Sobez2020, Molassembler2023} for all graph-theoretical algorithms. 
As part of this work, GP regression has been made freely available within our open-source \textsc{Scine}~\cite{scine} \textsc{Utilities} library.\cite{Utils2023}
 All timings were measured on a computer equipped with two Intel Xeon E5-2670 CPU with 2.60 GHz base frequency on one single thread.
Structures were visualized using \textsc{PyMol} 2.5.2.\cite{PyMOL}

\section{Automated Structure Preparation Pipeline ASAP}
\label{sec:asap}
The ASAP pipeline generates accurate structural models from some given structural input data (with the preferred file format being the PDB format).
The preparation is implemented as a sequence of three consecutive steps: In step 1, sequence analysis and error detection are carried out through a subgraph isomorphism ansatz that sorts out potentially faulty parts of the overall structure. In step 2, automated protonation is performed. In step 3, the protonated structure is solvated and the corresponding topology information for subsequent MM parametrization step is generated. 

\begin{figure}[h!]
    \centering
   \includegraphics[width=1.0\textwidth]{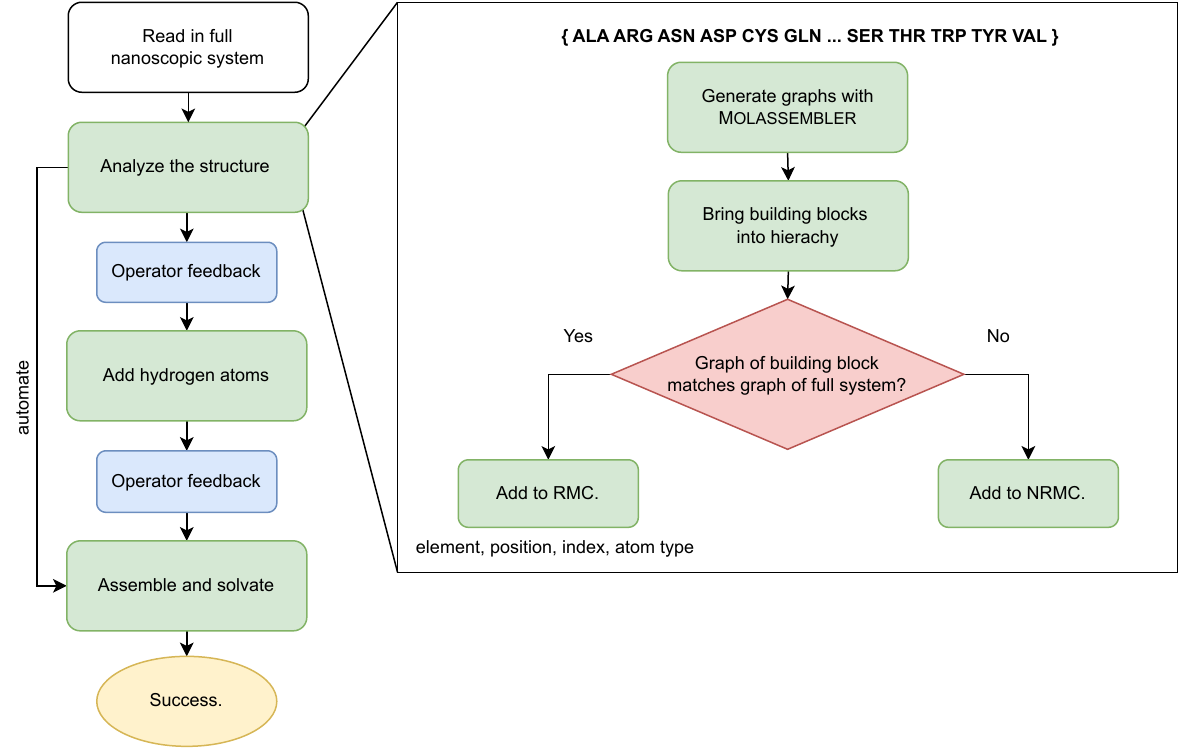}
    \caption{Schematic workflow of the ASAP pipeline (left) and the graph-based structure analysis step 1 (right) as implemented in \textsc{Scine Swoose} (see text for further explanation). The preparation steps are highlighted in green.}
    \label{fig:flowchart}
\end{figure}

As part of \textsc{Scine Swoose}\cite{swoose1.0.0}, we provide
both, a fully automated and a semi-automated variant of the ASAP pipeline, which offers high flexibility with regard to the degree of interference between operator and software. In the semi-automated supervised variant, the result of each step is provided as temporary structural output data. These data can be viewed, corrected, and fed back into the pipeline for subsequent steps. The workflow is shown in Figure \ref{fig:flowchart}.  

The building blocks for which the nanoscopic structure is searched must be sorted such that the largest building block is searched for first (to avoid false detection of occurrences in cases where smaller building blocks are elements of larger building blocks, \emph{i.e.}, glycine is a substructure of almost all other amino acids).  
We discuss the individual steps of ASAP in the following sections. For a detailed step-by-step description, we refer to the \textsc{Scine Swoose} manual on GitHub.

\subsection{Graph-Based Automated Structure Analysis}
\label{subsec:step_1}
In step 1 of the ASAP pipeline, a raw molecular structure is subjected to error detection. 
The principle idea of our structure analysis algorithm relies on the assumption that large parts of a nanoscopic structure often consist of recurring building blocks. Examples are proteinogenic amino acids in proteins, nucleotides in nucleic acids, triglycerides in lipid double layers, monosaccharides in oligosugar molecules. 

We start by automatically converting the molecular input structure to a graph employing the \textsc{Molassembler}\cite{molassembler1.2.1} library. Furthermore, all building blocks of a nanoscopic structure must be encoded as molecular graphs. Then, occurrences can be efficiently detected by application of graph-isomorphism techniques, that is, by querying the molecular graph of the overall structure for occurrences of each building block, which has been successfully demonstrated for related problems.\cite{Artymiuk1994,Spriggs2003} 
For the graph construction, we apply flexible edge (element type) and vertex (topography; in our case: distance-based connectivity) matching criteria, and do not evaluate three-dimensional features of the building blocks (stereochemistry, conformers, and so forth).
Note that we define a chemical bond based on the sum of the relevant van der Waals radii (including an additional threshold of 0.4 \r{A}).
We provide the graph information of each building block as string representations of the corresponding molecular graph. 

The graph-based approach offers the advantage of a flexible redefinition of the "building block", because any molecular 
structure can be converted to a molecular graph. In this work, our set of building blocks consists of the 20 proteinogenic amino acids, complemented by their selenium-containing analogs of methionine and cysteine (selenomethionine, selenocysteine), and of pyrrolysine. 

We fragment the input structure into 
a container for building blocks and into a container for all 
substructures that either consist of non--motif structures (containing non-amino-acid structures or amino acids with incorrect connectivity or missing atoms). We denote the resulting containers ``regular module container" (RMC) and ``nonregular module container" (NRMC), respectively. 
If applied to a protein structure, the RMC containers will collect all amino acids that form one or many polypeptide chains 
and the NRMC container will collect all non-amino acid 
structure subunits. 
This search algorithm efficiently characterizes the great majority of the input structure, while only a small remainder (the NRMC) 
is to be carefully checked for errors by the operator. This transfers the needle-in-a-haystack search, which comprises an exhaustive manual enumeration of the nanoscopic structure, to a tailored, system-specific and local evaluation of erroneous and nonstandard sites that are buried within a large structure. 
For instance, amino acids with missing atoms will be transferred to the NRMC, and can be corrected by the operator. This procedure is currently favored over automatic procedures in order to ensure a high fidelity of the structures.  
After finalization of step 1, a temporary output is generated, where the substructures are stored in separated files. 

Although currently optimized for protein-type nanoscopic systems, our 
software can in principle easily be extended towards other building blocks. For instance, a DNA strand
can be searched for occurrences of the nucleobases by storing string representations of adenine, cytosine, guanine, and thymine.
Another example would be the treatment of membrane proteins with ASAP, where our algorithms can be extended by incorporating string representations of lipid building blocks into ASAP, which are then automatically checked for in the large structure through the subgraph-isomorphism ansatz.

Moreover, we implemented two additional features into the pipeline: First, if the input structure is given in PDB format, 
multiple states can be separated from one another if identified through a marker in the file (for instance, different conformers or snapshots 
that are condensed in a single file). With this information at hand, one can either process all input structures separately or pick one or a few of them for further preprocessing. Note that we recommend discarding most or all solvent molecules in this step (except those that might contribute to chemical reactivity because they are in the vicinity of the active site), and reinsert them in step 3 through a more systematic explicit cluster-solvation approach (such as the one developed by us in Ref. \citenum{Simm2020}).

\begin{figure}[h!]
    \centering
    \includegraphics[width=1.0\textwidth]{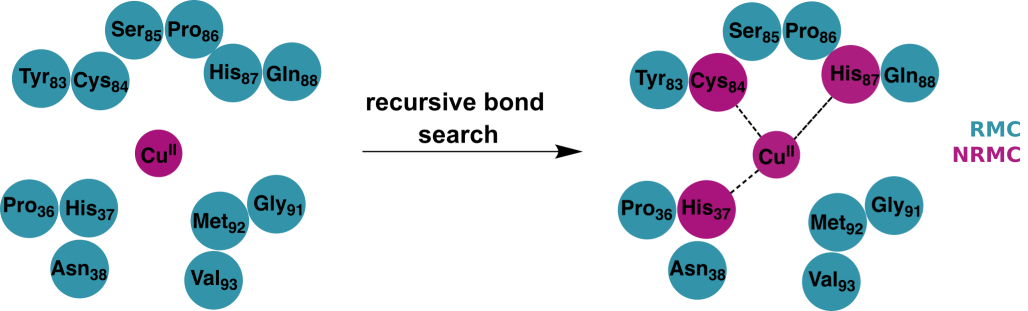}
    \caption{The recursive bond search algorithm implemented in the pre-processing  
    step of the structure preparation pipeline. }
    \label{fig:connectivity_plastocyanin}
\end{figure}

Second, we apply a recursive bond search strategy in which we iteratively transfer all correct building blocks that are covalently bound to at least one non-regular module atom (for instance, a metal center) also to the NRMC. 
This can be illustrated with a simple example: Structure processing of plastocyanin (PDB-ID: 1AG6~\cite{Xue1998}) separates the overall structure into 737 atoms in the regular-module polypeptide chain and one single $\mathrm{Cu^{2+}}$ atom stored in NRMC. After recursive bond search, the amino acids $\mathrm{Cys_{84}}$, $\mathrm{His_{37}}$, and $\mathrm{His_{87}}$ are transferred to NRMC, as shown in Figure \ref{fig:connectivity_plastocyanin}. 
The inclusion of the direct coordination environment around the metal site into NRMC
facilitates the assignment of charges and spin states, which is required to generate the SFAM model from QM reference data after structure preparation.

\subsection{Addition of Hydrogen Atoms}
\label{subsec:step_2_and_3}
\begin{figure}[h!]
    \centering
    \includegraphics[width=1.0\textwidth]{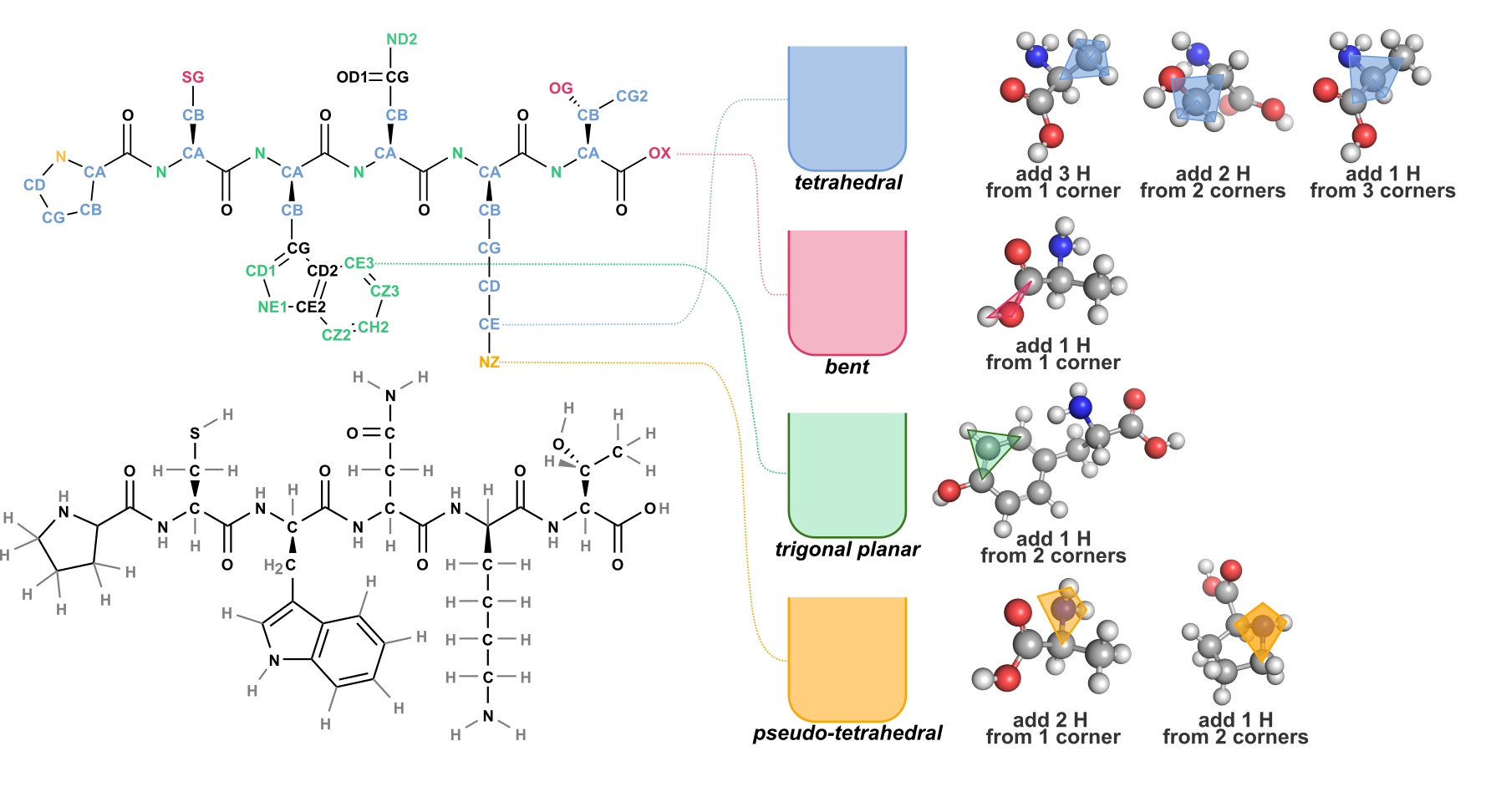}
    \caption{Hydrogen atom addition algorithm 
    for atom typing and grouping applied to a peptide chain of proline, cysteine, tryptophane, aspartic acid, lysine, and threonine as an example.}
    \label{fig:atom_typing}
\end{figure}

In step 2 of ASAP, hydrogen atoms (which are usually not resolved in crystallography) are added to both substructures in RMC and NRMC
separately. For this purpose, we exploit the redundancy of the recurring building-block structure motifs, 
which were fully characterized through the graph analysis in the previous step.
In an amino acid, the residue name and the atom type of each atom
together form a unique key from which the protonation pattern can be extracted. We apply the IUPAC-IUB rules for atom typing in amino acids~\cite{iupacAtomtyping}, which is the default format in PDB files.
In this format, atoms are labelled 
according to their distance from the carboxylate carbon atom; \emph{e.g.}, the directly bound C atom is denoted "CA". 
In phenylalanine (PHE), for instance, atom "CG" is a phenylic carbon atom, whereas in lysine (LYS), it is an aliphatic carbon atom, so that the protonation pattern for this atom type is only well-defined in combination with the respective amino acid.
With this information at hand, all atoms in the structure 
can be grouped according to 
how many hydrogen atoms must be added and in which geometric arrangement they should be placed (for the above example, PHE-CG is tetrahedrally coordinated, whereas LYS-CG is trigonal planar). This is schematically shown for
a peptide chain as an example in Figure \ref{fig:atom_typing}. For the 23 amino acids, the above criteria define 
\textit{tetrahedral} groups (with one to three hydrogen atoms to be added),
\textit{bent} groups 
(with one hydrogen atom added), \textit{trigonal planar} groups (with one hydrogen atom added) and \textit{pseudo-tetrahedral} groups (with one or two hydrogen atoms added and at least one polyhedral corner remains vacant). 
Assigning atoms into one of the above groups is trivial at first glance, but it can become complicated if the protonation state of a site depends on the chemical environment (that is, the pH value and the solvent).

The NRMC substructures are protonated by the external library \textsc{OpenBabel}~\cite{OpenBabel}. We emphasize that \textsc{OpenBabel} fills open valences with hydrogen atoms based on a hybridization assignment according to bond lengths and angles, so the result must be inspected with care. 
For this purpose, this step generates temporary output holding the protonated substructures in a human-readable format.

\subsection{Assembly and Solvation}
The third and last step in ASAP reassembles the final atomistic structure from the protonated RMC and NRMC substructures. Once merged, the substructure boundaries are cleaned up, because superfluous hydrogen atoms might have been added as an artifact of the fragmentation. Then, charge information are generated for the full structure. For the RMC, this is realized through application of simple chemical rules that derive atomic charges from the coordination environment of an atom. 
Charges and corresponding spin states for the NRMC need to be provided by the operator. 
Also, topological information about the system's connectivity
are generated, and the final structure is solvated by application of a systematic microsolvation approach 
developed in our group.\cite{Simm2020}
As a success criterion, we evaluate with \textsc{Swoose}
whether the SFAM MM model\cite{Brunken2020} can be generated for the resulting structure: During SFAM model parametrization, the atomistic structure is fragmented, and for each fragment, basic validity checks are carried out (\textit{e.g.}, it is analyzed whether the assigned charge and multiplicity form a fitting pair).  

While the RMC is handled automatically by ASAP and always delivers chemically correct structures (i.e., valences, 
charges, and multiplicities are reasonable and valid), errors in the assembled full system can only be introduced by the operator in the NRMC; for instance, by assigning incorrect charge-multiplicity pairs. For that reason, inspection and manipulation of the NRMC must be carried out carefully. Any introduced errors will later be detected during the parametrization of SFAM through aforementioned validity checks. 

\subsection{Machine Learning p$K_\mathrm{a}$\xspace Prediction from a Quantum Reference}

The assignment of a discrete protonation state for specific sites buried in a large structure as a function of environmental conditions (pH, solvent) is a long-standing problem.
As discussed above, the p$K_\mathrm{a}$\xspace can be predicted from quantities calculated with electronic structure methods, that is the free energy of deprotonation in solution.
The unfavorable scaling of these methods with the number of electrons, however, hampers p$K_\mathrm{a}$\xspace prediction through Equation (\ref{eq:pka}) for nanoscopic structures. 

\begin{figure}[h!]
    \centering
    \includegraphics[width=1.0\textwidth]{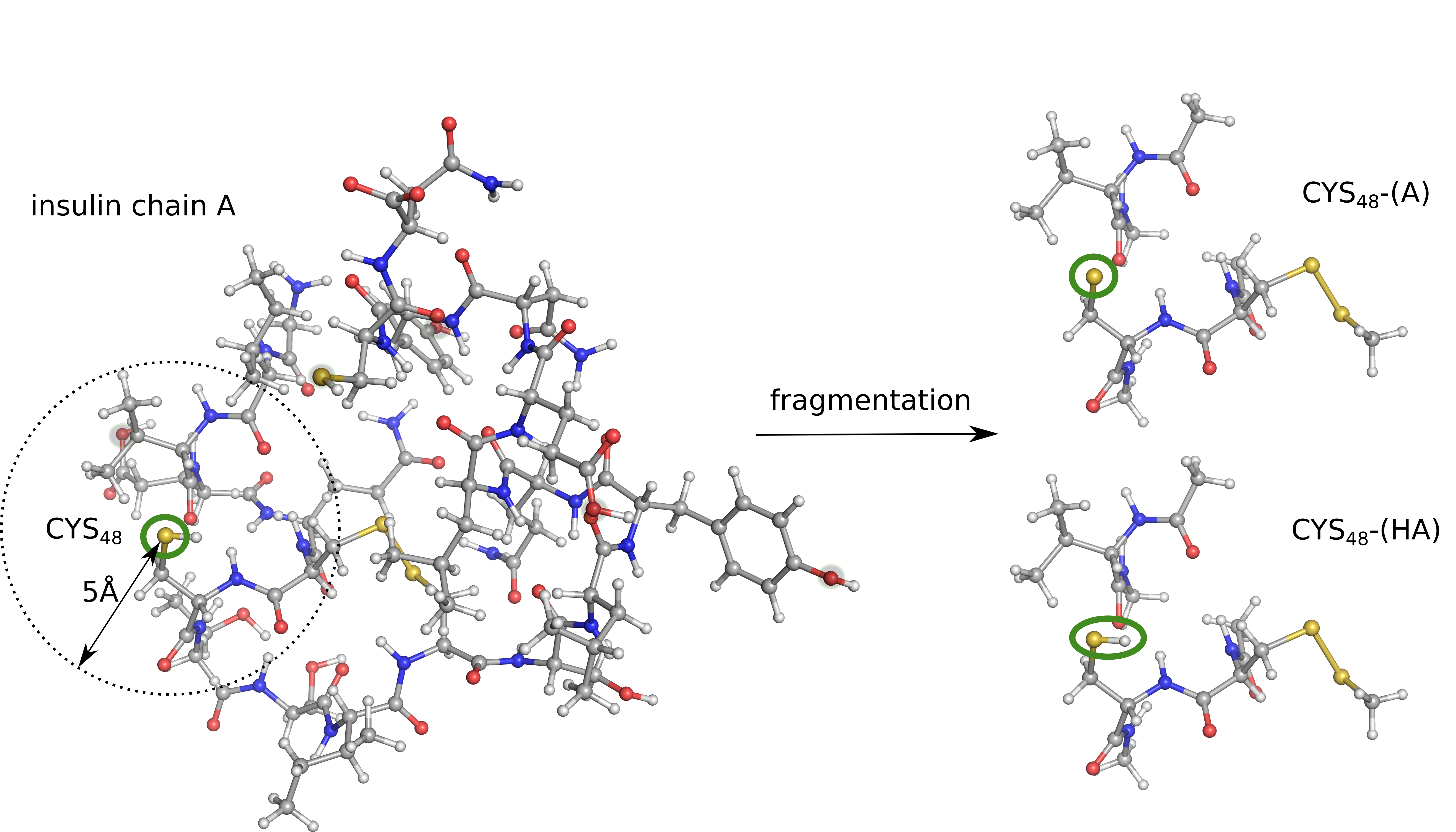}
    \caption{Fragmentation protocol applied to chain A of insulin, shown at the example of the pH-sensitive site in residue $\mathrm{CYS_{48}}$. The corresponding acid (HA) and base (A) fragments are generated. 
    Atom coloring: carbon, gray; hydrogen, white; oxygen, red; nitrogen, blue; and sulfur, yellow.}
    \label{fig:fragmentation_insulin}
\end{figure}

ASAP implements a parallelizable and physically consistent (i.e., derived from quantum mechanical calculations, typically DFT, instead of 
empirical rules)
p$K_\mathrm{a}$\xspace prediction protocol that is feasible for nanoscale structures by virtue of a divide-and-conquer strategy.
This is achieved by coupling the linear regression procedure reviewed in section \ref{subsec:pka_from_quantum_reference} to the fragmentation scheme developed in the context of the SFAM model parametrization.\cite{Brunken2020} The fragmentation scheme allows for the automated generation of spherical fragments around some central atom, which, in the present context, is a pH-sensitive atom. 
The radius of the sphere defining the size of the generated fragment should be chosen by two criteria: First, the QM reference calculations for the generated fragment must be computationally affordable. Second, the size of the fragment sets a cut-off radius that determines to what extent long-range electrostatic interactions, which affect 
the protonation state, are explicitly evaluated. Every interaction beyond this cut-off radius
will be disregarded, as it is not part of the fragment structure. For each site, both a protonated (AH, acidic) and a corresponding deprotonated 
(A, basic) fragment are automatically constructed, as is shown for one cysteine residue in chain A of insulin in Figure \ref{fig:fragmentation_insulin}. 

The two fragments are subjected to structure optimizations and subsequent Hessian calculations. These reference calculations are carried out during the MM model parametrization step and controlled by the \textsc{Scine} database module~\cite{database1.0.0}. This procedure offers several advantages: First, it exploits the efficient parallelization of the MM model parametrization at negligible additional cost. For example, to generate reference data for the SFAM model of insulin chain A, which consists of 328 atoms, 328 structure optimizations, and 328 subsequent Hessian calculations need to be performed, because a fragment is generated around each atom (the resulting redundancy can be exploited for internal consistency checks). The peptide chain contains only six pH-sensitive sites, that is, 12 additional reference calculations need to be performed for the corresponding second state of the site. Also, the force constants required for MM parametrization can be extracted from these reference calculations for both species. Information about them enables a change of the  pH-value during a molecular simulation 
with fast reparametrization of SFAM without the need to generate additional reference data for a different protonation state. 
In transferable force fields, the parameters can easily be adjusted as the atom type changes upon addition or removal of a proton. In the context of system-focused atomistic models, however, the force field parameters have been optimized for a specific molecular configuration, so that an efficient strategy for parameter re-optimization will be needed if minimal structural rearrangements occur.

\section{Results and Discussion}
\label{sec:results}
\subsection{Automated Structure Processing at the Example of Insulin Chain A}
\label{subsec:automated_processing_insulin}
\begin{figure}[h!]
    \centering
    \includegraphics[width=1.0\textwidth]{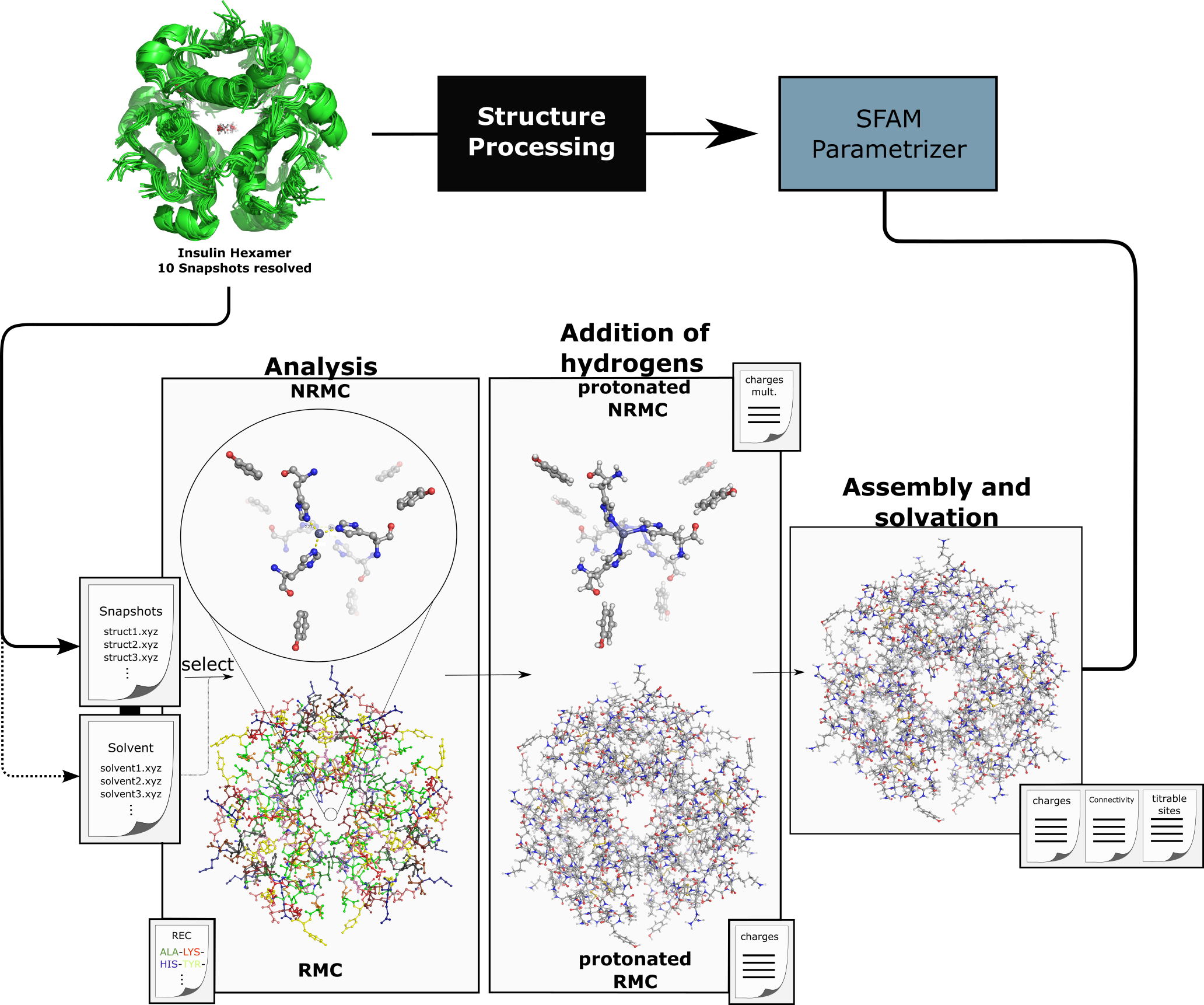}
    \caption{Schematic representation of the ASAP pipeline applied to chain A of insulin.} 
    \label{fig:processing_insulin}
\end{figure}

We selected chain A of the peptide hormone insulin as an example to demonstrate the capabilities of  ASAP. The overall workflow, the resulting structures, and the generated output information are compiled in Figure \ref{fig:processing_insulin}. The starting structure was a hexameric hormone that featured 10 conformers resolved in the experimental PDB file and was taken from PDB-ID 1AI0.\cite{Chang1997}
The zinc active site ($\mathrm{Zn^{2+}}$) coordinates three histidine ligands and one water ligand in a tetrahedral configuration.

We applied step 1 of ASAP to separate the overall structure into its 10 conformers, all being candidates for further preprocessing. In all conformers, 
only two solvent molecules were resolved in the structure close to the active site, and they were therefore kept. After separation of the 10 candidates, we selected candidate 4 for further preprocessing. In all other candidates, 
at least two of the three histidine ligands were
coordinated to the metal site at a distance of at least 2.4 \r{A}, 
which clearly exceeds both the tight range of 1.95-2.10 \r{A} for this type of bonds known from experiment\cite{Zhu1997} and our definition of a chemical bond derived from van der Waals radii (see above).
Note that none of the 10 candidate structures exhibited interatomic distances that fit either one of the two criteria for all Zn-coordinating ligands.

The graph-based analysis in step 1 resulted in a separation into 2380 RMC atoms and 96 NRMC atoms, as depicted in the first gray box in Figure \ref{fig:processing_insulin}. 
The NRMC container was filled with the metal-center active site and the corresponding ligands. In the second step, 
hydrogen atoms were added to the RMC substructure
by applying the clustering strategy based on atom typing and amino acid categorization as described above. All pH-sensitive sites were protonated such that the site is uncharged, by default.
After the addition of hydrogen atoms to the NRMC substructure, we assigned a charge of $+2$ to both zinc atoms (central gray box in Figure \ref{fig:processing_insulin}).
Ultimately, the final processing step 3 assembled both 
substructures and corrected the boundaries (because superfluous hydrogen atoms were added as a result of cutting bonds at the substructure interface), 
and determined the topology by a distance-based connectivity evaluation. All corresponding information was stored in separate files. 
This yielded the final structure depicted in the right gray box in Figure \ref{fig:processing_insulin}).

\subsection{ASAP Preparation Benchmark and Performance Evaluation}
To investigate the overall efficiency and stability, we applied 
ASAP to three different data sets: The RS126 data set~\cite{Rost1993}, the CB396 data set~\cite{Cuff1999} and the EVA6~\cite{Rost2001} data set. The corresponding PDB identification codes are compiled in Table S1 of the Supporting Information.  
We combined these three data sets to one large test set, which 
consists of, in total, 589 structures. 
After ASAP processing, 
we found that only 210 structures required no further preprocessing, which corresponds to 35 \% of the data set.
That is, they consisted exclusively of amino-acid motifs that have a chemically correct composition and connectivity. Overall, 153 structures have at least two conformations resolved. 
For 88 $\%$ of the test-set compounds (518 structures), we found that the NRMC substructure accounted for less than 5 $\%$ in terms of atom count of the full system. Hence, in most cases, only a very small portion of the full structure requires manual inspection and curation.

For completeness, we compared the results obtained with ASAP for selected PDB-IDs to the \textsc{PdbFixer} software provided by OpenMM. \textsc{PdbFixer} also realizes different degrees of automation and precise control of every part of the process. The main difference to ASAP is that the pH-specific protonation states are derived empirically, and that it cannot be directly coupled to the parametrization of a system-focused atomistic model. 

\begin{figure}[h!]
\begin{minipage}[t][][b]{0.48\textwidth}
\centering
\includegraphics[width=1.0\linewidth]{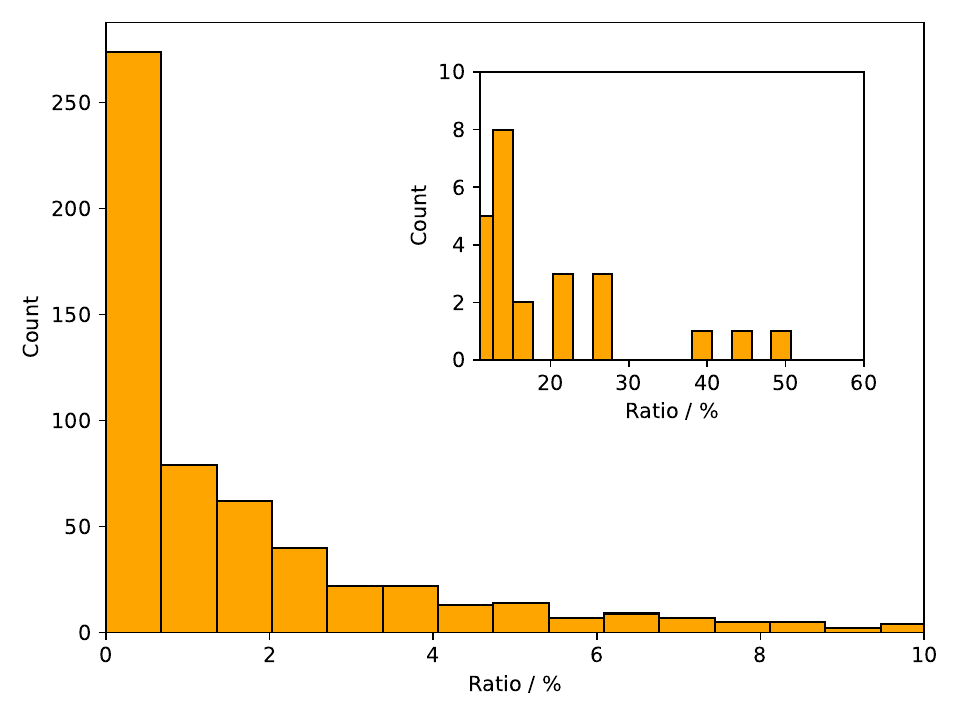}
\caption{(i) Atom-count ratios between REC and NREC substructures for the test sets.}
\label{fig:prob1_6_2}
\end{minipage}
\hfill
\begin{minipage}[t][][b]{0.48\textwidth}
    \centering
    \includegraphics[width=1.0\linewidth]{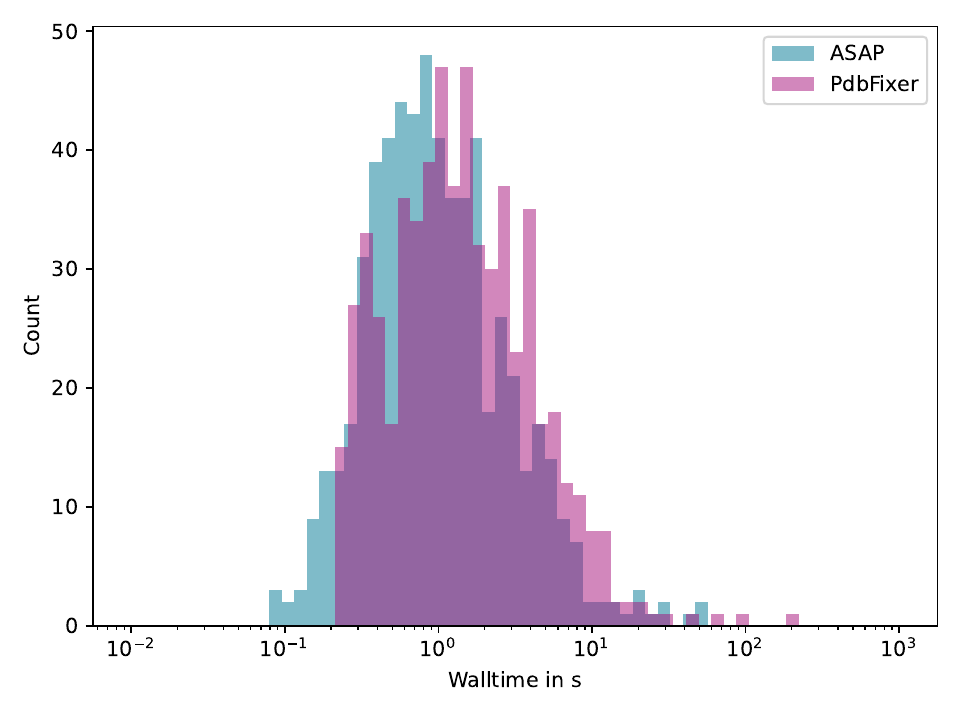}
    \caption{Distribution of wall times for the preparation of the CB396, RS126, and EVA6 datasets with \textsc{PdbFixer} (purple) and \textsc{Asap} (blue). The \textsc{Asap} data refer to the graph-based analysis step only. }
    \label{fig:runtime-histogram-pdbfixer-vs-asap-prepare-analyze}
\end{minipage}
\end{figure}

Regarding performance, our focus in this discussion is primarily on assessing the time required for the graph isomorphism step. This is driven by two key factors. 
First, it is the most time-consuming step in ASAP. Second, a majority of the structures in the test set is unsuitable for fully automated processing because they contain multiple non-standard atoms or residues, as illustrated in Figure \ref{fig:prob1_6_2}. Therefore, an analysis of the times of the fully automated preparation step would not be representative of the philosophy of ASAP. The results are summarized in Figures \ref{fig:prob1_6_2} and \ref{fig:runtime-histogram-pdbfixer-vs-asap-prepare-analyze}.

We emphasize that structure preparation is computationally negligible compared to subsequent SFAM parametrization, which can consume days to weeks for large systems. 
For the largest error-free structure (i.e., a structure that can be processed in full automation), all three steps were completed within few minutes. 

\subsection{Quality Assessment of Structures prepared with ASAP}

To assess the structural integrity and validity of structures generated by ASAP, we subjected a subset of the dataset to short MD simulations (0.3 $ns$ pressure equilibration and 1 $ns$ NPT production run), and evaluated selected features from these molecular dynamics trajectories. We furthermore compared the standard deviation of the total energy as a criterion for overall structural stability to results from structures 
processed with \textsc{PdbFixer} of OpenMM\cite{openMM}. Currently, our system-focused atomistic model is not fully compatible with available MD software. This is due to the fact that we generate system-focused atom-types and force field parameters for every individual structure, together 
with customized functional forms (see Ref. \citenum{Brunken2020}), which can not directly be interfaced to a standard MD engine. Therefore, we only selected those structures in the dataset that solely consist of RMC substructures (which are, in total, 243 structures of our dataset). In this case, the ASAP-structures (provided as \textsc{XYZ} coordinates) can be converted into PDB files that are compatible with transferable force fields. For all MD calculations, we employed the \textsc{OpenMM} program package and the ff14SB force field\cite{ff14sb}. The pressure was set to 1 $bar$, the temperature to 298.25 K, the time step to 2.0 $fs$, the friction to 1.0 $ps^{-1}$, and every 100th step being recorded. All respective structures have been processed by ASAP in a fully automated manner, i.e., no operator interference was required, but the basic ASAP suggestions were applied.

\begin{figure}[h!]
    \centering
    \includegraphics[width=1.0\textwidth]{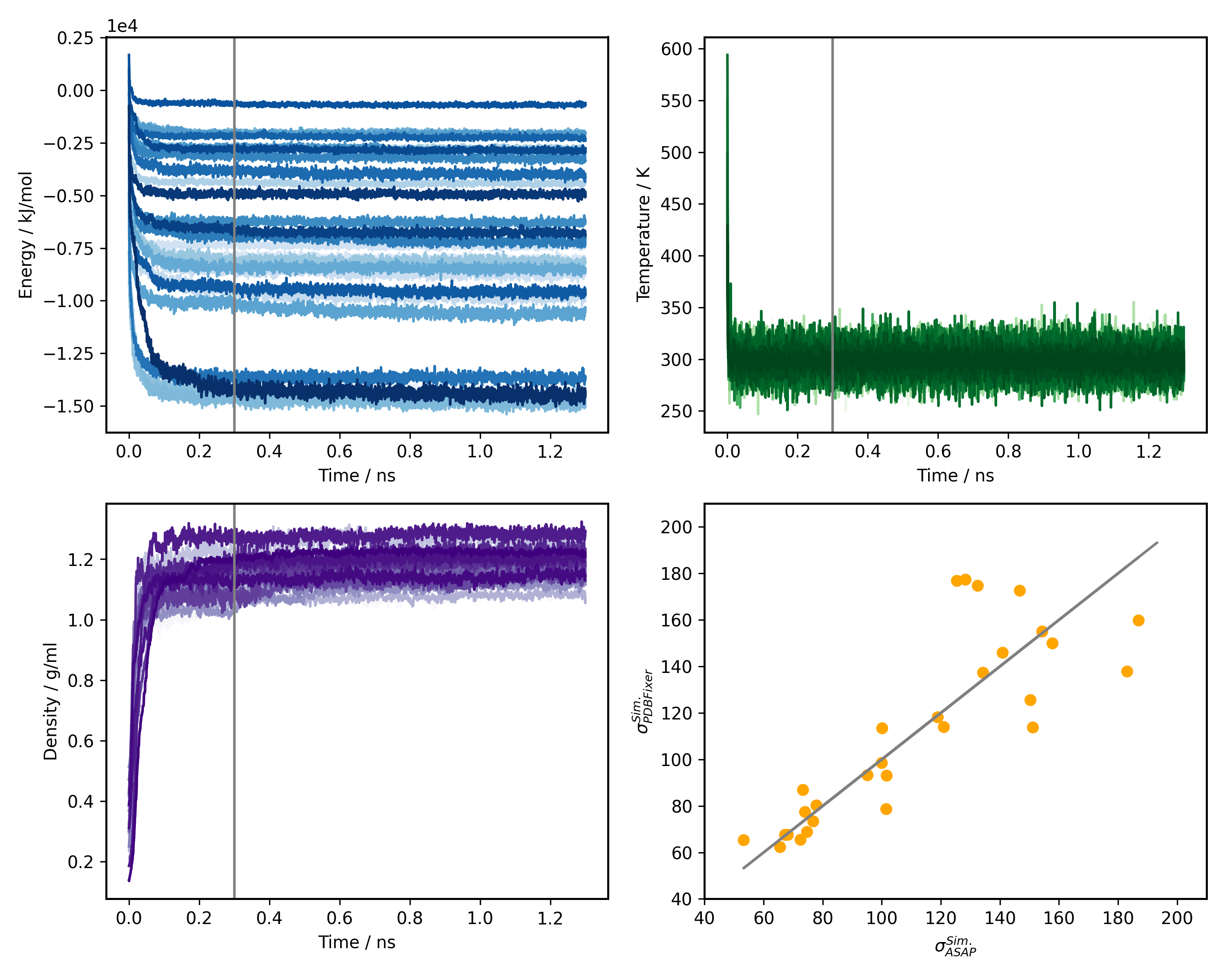}
    \caption{Total energy (top left), temperature (top right), and density (bottom left) as a function of simulation time for selected structures from the test set. The gray line indicates the end of NPT equilibration and the start of the actual simulation. On the bottom right, the standard deviation of total energy during the simulation (0.3 ns $\geq$ t $\geq$ 1.3 ns) of ASAP-structures compared to \textsc{PdbFixer}-structures is depicted. The respective structures and trajectories generated by ASAP and the \textsc{PdbFixer} can be found in the ASAPdata dataset on Zenodo.}
    \label{fig:asap-stability-assessment}
\end{figure}

As illustrated in Figure \ref{fig:asap-stability-assessment}, all inspected ASAP-structures remain stable over a simulation time of 1 ns, maintaining structural integrity. Furthermore, density and temperature quickly converge, which is indicative of effective calibration and high-quality structures. The standard deviation of the total energy during the simulation is on par or similar to the well-established \textsc{PdbFixer}.

\subsection{Automated p$K_\mathrm{a}$\xspace Prediction}
\subsubsection{Construction of a Suitable Training Set}

\begin{figure}[h!]
    \centering
    \includegraphics[width=1.0\textwidth]{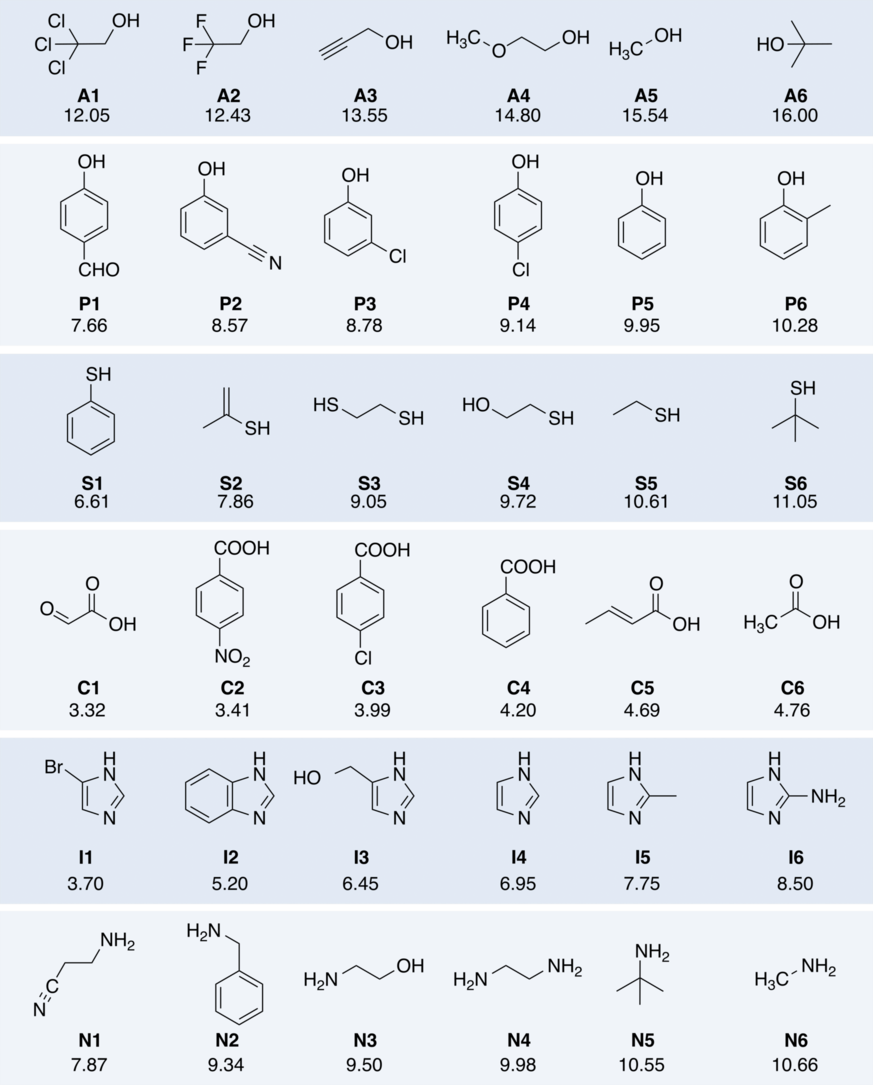}
    \caption{Training-set molecules with known experimental $\mathrm{pK_a}$\xspace values (given below the label for the molecule)
    that were employed in the GP. The compounds are labeled according to their functional group: Alcohol groups: \textbf{A1-A6}, phenol groups: \textbf{P1-P6}, thiol groups: \textbf{S1-S6}, carboxylic acid groups: \textbf{C1-C6}, imidazole groups: \textbf{I1-I6}, amines: \textbf{N1-N6}.}
    \label{fig:gpr_training_molecules}
\end{figure}

To train the GP model, we assemble six small-molecule data sets for each of the six different
pH-sensitive functional groups that can be found in amino acids (\emph{vide infra}).
These groups 
comprise thiol groups (--SH), alcohol groups  (--OH), phenol groups (Ph--OH), carboxylate groups (--COOH), imidazole groups (Im-H), and primary amines (--NH2).
Once ${\Delta G_0^\mathrm{aq}}$ is calculated for all training-set molecules, the GP must be trained only once for each combination of functional-group class and electronic structure method. 
We selected candidate molecules for the six training sets 
according to two criteria: (i) Accurate experimentally measured p$K_\mathrm{a}$\xspace values should be available for
(ii) preferably small and rigid molecular structure. 
The second criterion mitigates the computational cost for training data evaluation
for different conformers. The training-set molecules and corresponding p$K_\mathrm{a}$\xspace values for phenol and alcohol groups were taken from Ref. \citenum{Thapa2017a}, for thiol groups from Ref. \citenum{Thapa2016}, for imidazole groups from Refs. \citenum{SchmidtAmBusch2004} and \citenum{Matsui2012}, for amines from Refs. \citenum{Matsui2012} and \citenum{Juranic2014}, and for carboxylate groups from Ref. \citenum{Beale2013}; see Figure \ref{fig:gpr_training_molecules}. Their Cartesian coordinates are given in Tables S2$-$S7 in the Supporting Information.

\begin{figure}[h!]
    \centering
    \includegraphics[width=1.0\textwidth]{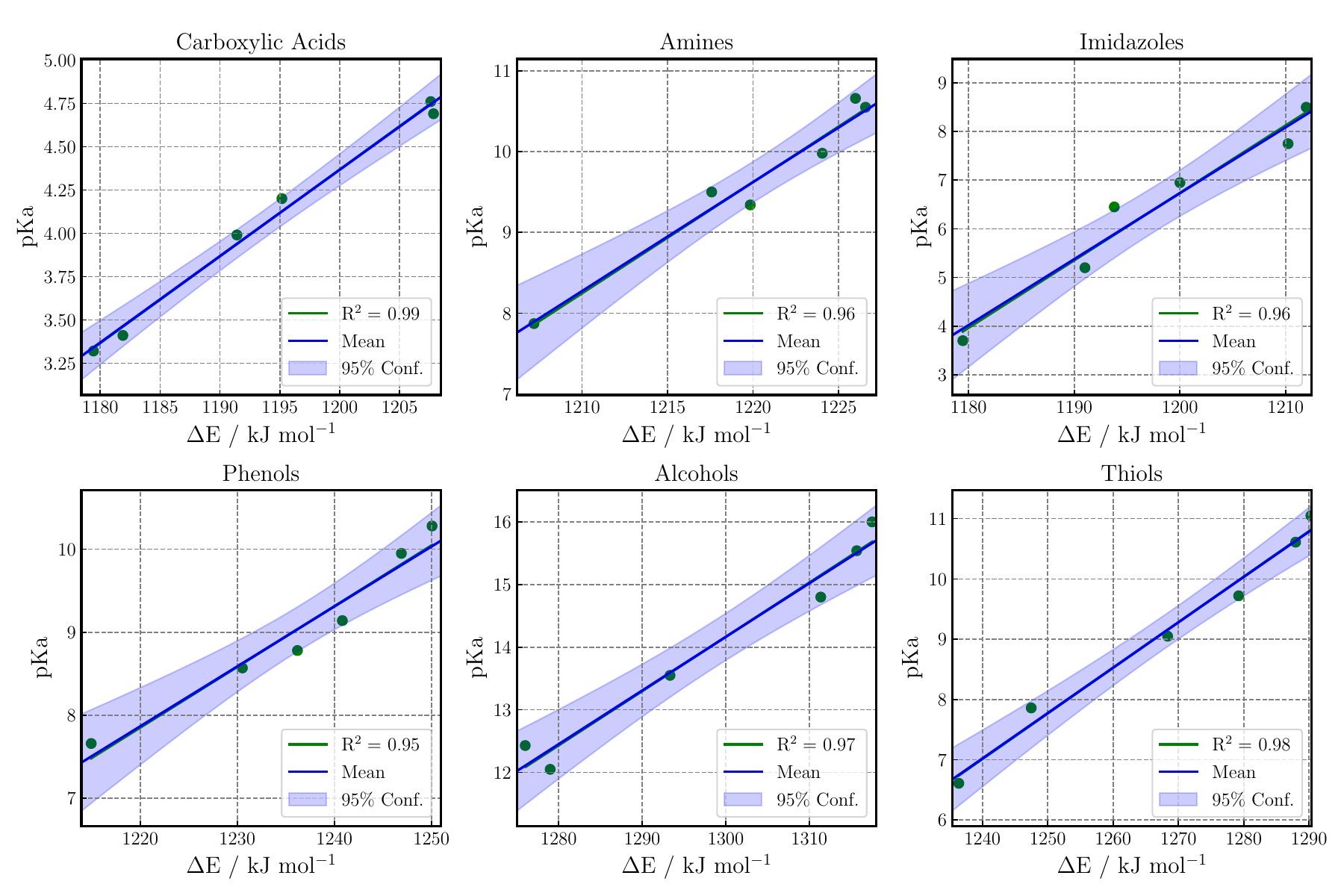}
    \caption{GP regression (blue) and corresponding linear regression (green) of training data from Figure \ref{fig:gpr_training_molecules}. }
    \label{fig:training_molecules_regression}
\end{figure}

We evaluate the relationship between calculated ${\Delta G_0^\mathrm{aq}}$ and experimental reference p$K_\mathrm{a}$\xspace through comparison of ordinary linear regression and GP regression. As shown in Figure \ref{fig:training_molecules_regression}, the GP efficiently learns the linear relationship from only six data points each. Compared to linear regression, which assumes a homoscedastically distributed error, we find that the uncertainty of the predicted p$K_\mathrm{a}$\xspace from the GP is higher in regions where our training-set data are sparse (and grows even larger outside the boundaries of the training set, which is omitted in Figure \ref{fig:training_molecules_regression}). GP models work well for the given regression problem, because they can be trained on small training data sets. However, the training data presented in this work may be extended to cover an even broader pH region.

We investigated whether semiempirical methods are a viable alternative to conventional DFT, 
as they allow for very fast p$K_\mathrm{a}$\xspace prediction even for large structures at moderate computational cost and have therefore been widely applied to nanoscopic structures.\cite{Kromann2016, Jensen2017, Hunt2020} 
The overall findings indicate that the reliability of semiempirical methods in predicting protonation probabilities should be carefully assessed. For a detailed discussion, see Section SIV in the Supporting Information. 

\clearpage

\subsubsection{p$K_\mathrm{a}$\xspace Prediction For Isolated Amino Acids in Aqueous Solution}
We applied our trained GP model to report p$K_\mathrm{a}$\xspace values of the isolated amino acids glutamic acid (GLU), aspartic acid (ASP), lysine (LYS), tyrosine (TYR), histidine (HIS) and cysteine (CYS) because there are reference p$K_\mathrm{a}$\xspace values available for these amino acids at standard conditions in aqueous solution. We calculated p$K_\mathrm{a}$\xspace values in two ways: 1) free energy difference of acid and base in solution were evaluated according to Equation (\ref{eq:pka_from_linear_relationship}), and 2) electronic energy differences, which do not contain entropy and enthalpy contributions of the nuclear scaffold in the condensed phase, were considered as an efficient 
approximation.
The predicted p$K_\mathrm{a}$\xspace values are compiled in Table \ref{tab:pka_amino_acids}.

\begin{table}[h!]
\centering
\label{tab:pka_amino_acids}
\caption{Predicted p$K_\mathrm{a}$\xspace values of side chains of different amino acids (AA) obtained with GP regression based on PBE-D3 and PBE0-D3 
results compared to model p$K_\mathrm{a}$\xspace~\cite{Fitch2002, Thurlkill2005} data at standard conditions. Mean absolute errors (MAE) were calculated as the sum of the absolute errors between the reference p$K_\mathrm{a}$\xspace values of the six amino acids, and ASAP's predictive mean, weighted equally.
For each density functional, the first column contains p$K_\mathrm{a}$\xspace values learned from electronic energy differences of acid and base ($\mathrm{\Delta E_0^\mathrm{aq}}$), while the second column derives this property from Gibbs free energy differences ($\mathrm{\Delta G_0^\mathrm{aq}}$). For an overview of p$K_\mathrm{a}$\xspace values of amino acids in proteins, we refer the interested reader to Ref. \citenum{Thurlkill2005} and references therein.}
\begin{tabular}{lcrrcrr}
\hline
\hline
\multicolumn{1}{c}{\multirow{3}{*}{AA}} &\multicolumn{1}{c}{\multirow{3}{*}{  Ref. p$K_\mathrm{a}$}} & \multicolumn{5}{c}{predicted p$K_\mathrm{a}$\xspace value}  \\
\cline{3-7}
& & \multicolumn{2}{c}{PBE-D3} & &\multicolumn{2}{c}{PBE0-D3}  \\
& & from ${\Delta E_0^\mathrm{aq}}$ & from ${\Delta G_0^\mathrm{aq}}$ &  &from ${\Delta E_0^\mathrm{aq}}$ & from ${\Delta G_0^\mathrm{aq}}$ \\
\hline
GLU        &     4.4         & 4.20 $\pm$ 0.26   & 4.74 $\pm$ 0.41 
 &                            & 4.67 $\pm $ 0.38  & 4.86 $\pm$ 0.45 \\
 
ASP         &    4.0         & 4.37 $\pm$ 0.29   & 4.42 $\pm$ 0.31 
 &                            & 4.39 $\pm$ 0.29   & 4.53 $\pm$ 0.34 \\
 
HIS        &     6.3         & 6.52 $\pm$ 0.25   & 6.54 $\pm$ 0.25 
 &                            & 6.04 $\pm$ 0.26   & 6.18 $\pm$ 0.26 \\
 
CYS        &     8.5        & 8.38 $\pm$ 0.29  & 8.47 $\pm$ 0.28 
 &                            & 8.28 $\pm$ 0.30 & 8.49 $\pm$ 0.28 \\

TYR        &     9.6        & 9.73 $\pm$ 0.32   & 9.74 $\pm$ 0.32
 &                            & 9.58 $\pm$ 0.29   & 9.55 $\pm$ 0.29 \\
 
LYS        &     10.4       & 10.07 $\pm$ 0.28  & 10.43 $\pm$ 0.33 
&                             & 9.42 $\pm$ 0.26   & 10.32 $\pm$ 0.32 \\
\hline
\hline
\textbf{MAE} & & \multicolumn{1}{c}{0.22} & \multicolumn{1}{c}{0.26} & & \multicolumn{1}{c}{0.36} & \multicolumn{1}{c}{0.20} \\
\hline
\end{tabular}
\end{table}

We found that the predicted p$K_\mathrm{a}$\xspace values for the six amino acids are overall consistent with the experimental values, with an MAE of 0.30 p$K_\mathrm{a}$\xspace units for PBE-D3 and 0.43 p$K_\mathrm{a}$\xspace units for PBE0-D3.
The PBE0-D3 functional slightly overestimated the p$K_\mathrm{a}$\xspace values compared to the PBE-D3 functional with a mean signed error of 0.19 p$K_\mathrm{a}$\xspace units compared to 0.01 p$K_\mathrm{a}$\xspace units. Note that we found basis sets of at least triple-zeta quality to be required for sufficiently reliable predictions. In addition, we found no significant improvement in predicted p$K_\mathrm{a}$\xspace values when free energy differences were evaluated compared to electronic energy differences 
(for a demonstration of quantitative agreement in predicted p$K_\mathrm{a}$\xspace values see Table \ref{tab:pka_amino_acids}, and for the verification of a linear relation between $\Delta G$ and the p$K_\mathrm{a}$\xspace see Section III in the Supporting Information).
Hence, in the following, we chose for p$K_\mathrm{a}$\xspace prediction in nanoscopic structures electronic energy differences only. 

\subsubsection{p$K_\mathrm{a}$\xspace Prediction in Nanoscopic Structures} 
\label{subsec:pka_nanoscale}

\begin{figure}[h!]
    \centering
    \includegraphics{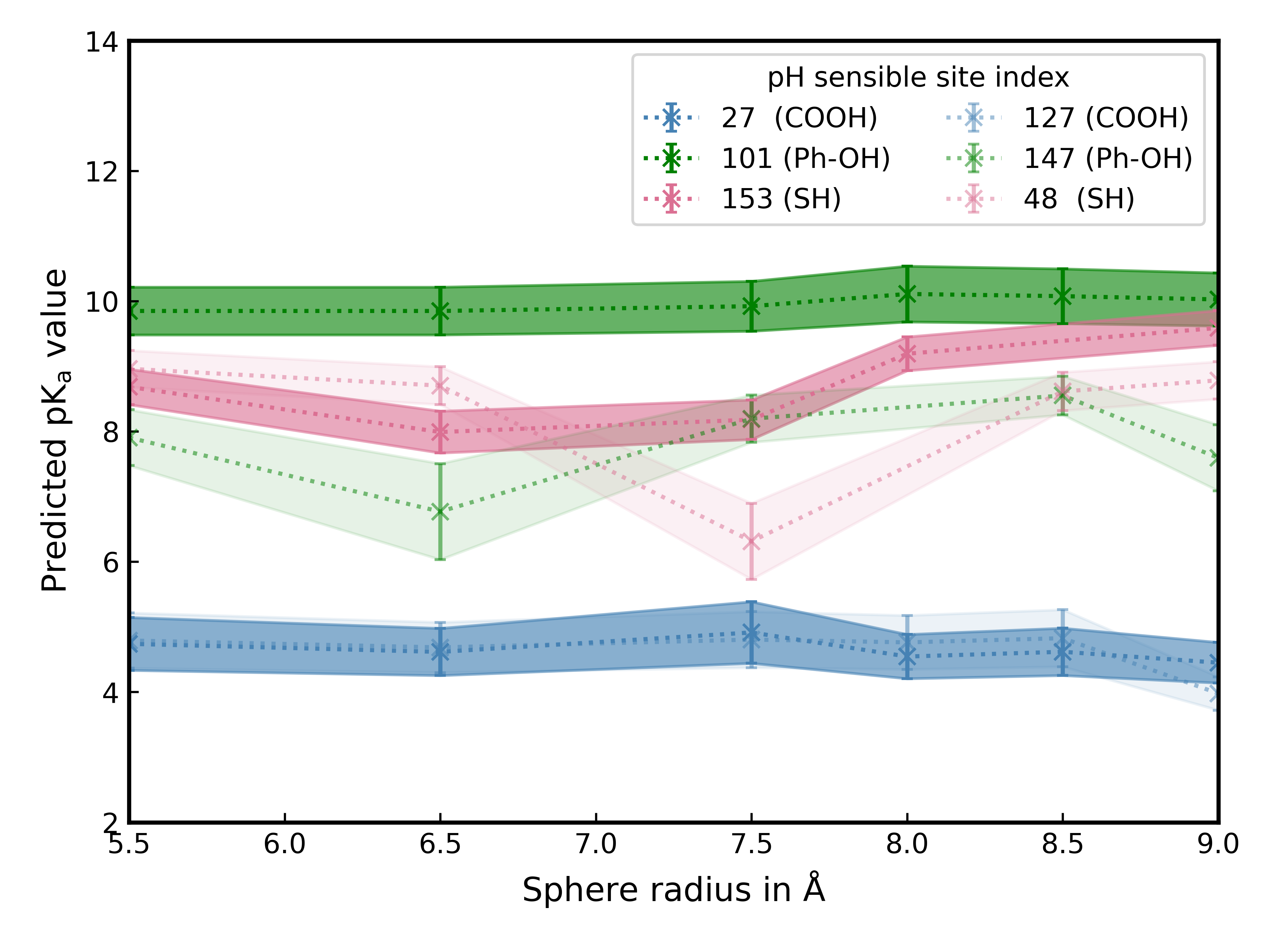}
\caption{Calculated p$K_\mathrm{a}$\xspace values for pH-sensitive sites in insulin chain A and the corresponding prediction uncertainty. Note that the lines between the data points should only guide the eye. Carboxylate 
groups are labelled in blue, phenol groups are labelled in green, thiol groups are labelled in pink. 
}
    \label{fig:pka_from_spheres_insulin}
\end{figure}

We demonstrate p$K_\mathrm{a}$\xspace prediction for nanoscale structures at the example of chain A of the peptide hormone insulin. The initial structure was taken from PDB-ID 1AI0.\cite{Chang1997}  
We calculated the p$K_\mathrm{a}$\xspace value of six pH-sensitive sites in insulin chain A, namely $\mathrm{GLU_{27}}$, $\mathrm{CYS_{48}}$, $\mathrm{TYR_{101}}$, $\mathrm{GLU_{127}}$, $\mathrm{TYR_{147}}$ and $\mathrm{CYS_{153}}$. The subscripts denote the index of the pH-sensitive atoms that were either protonated or deprotonated. As structure optimizations are very costly for structures with more than 300 atoms, we generated different-sized fragments around the pH-sensitive atoms and optimized the structures with PBE-D3/def2-TZVP. The size of the fragment determined the degree to which long-range electrostatic interactions were explicitly considered when evaluating the p$K_\mathrm{a}$\xspace value. 
We report the sensitivity of predicted p$K_\mathrm{a}$\xspace values with respect to system size in Figure \ref{fig:pka_from_spheres_insulin}.

We found that especially carboxylate groups (atom indices 
27 and 127 in the full structure) were very stable over a broad range of differently sized spheres, highlighting that all relevant interactions were accurately modelled even within the smallest sphere of 5.5 \r{A}. 
Moreover, we found that one of the phenolic pH-sensitive atoms (atom index 101) was insensitive toward fragment size, and one (atom index 147) exhibited slightly increased sensitivity.
For thiol groups, we found the largest variation with regard to system size and no convergence of the p$K_\mathrm{a}$\xspace value with increasing fragment size. 
This observation suggests that the variations in predicted acidity constants are not a consequence of lacking electrostatic interactions, but rather a deficiency of the method or the structure. 
This issue may be mitigated by application of explicit 
solvent models, more accurate basis sets, and the generation of multiple conformers. 

In the current implementation of ASAP, we do neither distinguish between the two protonation states of the two oxygen atoms in the carboxylic group of ASP and GLU, nor between those of the two nitrogen atoms in the imidazole group of HIS. This is because we assess protonation probabilities only after structure optimization, during which the R-C bond of R-COOH and R-Imidazole are not constrained, but in principle allowed to rotate, rendering the respective sites indistinguishable in principle. We stress that this rotation can very well be sterically constrained by the protein environment and would therefore require $2^{n}$ reference calculations for $n$ potential protonation sites. The same holds true for pH-sensitive sites of different amino acids that are spatially close, and can couple (as, for instance, GLU35 and ASP52 in HEWL\cite{Huang2016a}). These coupling of protonation states is currently not supported by ASAP.

Furthermore, we compared the results obtained with ASAP to those obtained with the software packages \textsc{Pypka} and \textsc{Propka} for $\mathrm{pK_a}$\xspace  prediction.
The results are summarized in Table \ref{tab:predicted_pka_different_methods}.

\begin{table}[H]
	\centering	\caption{\label{tab:predicted_pka_different_methods}$\mathrm{pK_a}$\xspace values in chain A of insulin predicted with different available models. The results from our model refer to fragments with an initial radius of 5.5\;\r{A}. Note that neither with \textsc{Pypka} nor \textsc{Propka} delivered $\mathrm{pK_a}$\xspace values for cysteine groups.}
	\begin{tabular}{cS[table-format=2.2]S[table-format=2.2]cS[table-format=2.2]}
 		\hline
   		\hline
		Amino Acid (AA) & \textsc{Pypka}\cite{Reis2020} & \textsc{Propka}\cite{Olsson2011}  & \textsc{Scine Swoose} & {isolated AA}\\
		\hline
		GLU$_{27}$     & 4.09  & 4.56 &  4.74 $\pm$ 0.41 & 4.4    \\
		CYS$_{48}$  & {--} & {--} & 8.96 $\pm$ 0.28 & 8.5    \\
		TYR$_{101}$   & 10.16  & 10.05  & 9.85 $\pm$ 0.37 & 9.6 \\
		GLU$_{127}$  & 4.58  &  4.66  & 4.79 $\pm$ 0.42  & 4.4    \\
		TYR$_{147}$   &  10.63 & 10.22   & 7.90 $\pm$ 0.43  & 9.6   \\
		CYS$_{153}$   &  {--} & {--}  & 8.68 $\pm$ 0.27  &     8.5         \\	
  \hline
  \hline 
	\end{tabular}
\end{table}

We find that the ansatz developed by us predicts $\mathrm{pK_a}$\xspace values in the same order of magnitude as the empirical approaches \textsc{pypKa} and \textsc{Propka} for all functional groups except for TYR$_{147}$. Here, our model estimates a $\mathrm{pK_a}$\xspace value of 7.90\;$\pm$\;0.43, whereas the other models predict  $\mathrm{pK_a}$\xspace values close to the reference value of the isolated amino acid tyrosine. The decrease of 3 $\mathrm{pK_a}$ units predicted by our model is due to a stabilization of the deprotonated state through a hydrogen bond with a neighboring amino-group, which increases the acidity of this site.

To further validate the quality of $\mathrm{pK_a}$\xspace prediction, we assess the predictive performance of ASAP for selected active sites in the Barnase structure (PDB-ID: 1A2P)\cite{Martin1996}), for which experimentally well-resolved $\mathrm{pK_a}$\xspace values are available in the literature. The results are summarized in Table \ref{tab:pka_comparison_experiment}.

\begin{table}[h!]
    \centering
    \caption{$\mathrm{pK_a}$\xspace values of selected residues in PDB 1A2P compared to experimental $\mathrm{pK_a}$\xspace values from Refs. \citenum{Oliveberg1995}, \citenum{Loewenthal1991} taken from the PKAD database\cite{pkad}. }
    \begin{tabular}{lcccc}
    		\hline
      		\hline
    Residue & Experiment &  ASAP & \textsc{Propka} & \textsc{Pypka} \\
    \hline
    ASP$_8$ 48  & 3.1 $\pm$ 0.1 \cite{Oliveberg1995}   &3.0 $\pm$ 0.6 & 2.86 & 2.9\\
    HIS$_{18}$ 130 & 7.8 $\pm$ 0.02 \cite{Oliveberg1995}   & 7.2 $\pm$ 0.3 & 5.37 & 6.79 \\
    GLU$_{73}$  526 &  2.1 $\pm$ 0.1 \cite{Loewenthal1991}  & 2.0 $\pm$ 1.7 & 5.18 & 2.73 \\ 
    HIS$_{102}$ 770   &  6.3 $\pm$ 0.05 \cite{Loewenthal1991} & 6.2 $\pm$ 0.3 & 5.87 & 6.11 \\
\hline
		\hline
    \end{tabular}
    \label{tab:pka_comparison_experiment}
\end{table}

For the selected pH-sensitive sites in 1A2P, we find excellent agreement between 
$\mathrm{pK_a}$\xspace values from experiment and predicted with ASAP. 
For the two histidine residues, 
ASAP data correctly reflect that the $\mathrm{pK_a}$\xspace of $\mathrm{HIS_{18}}$ is more than one magnitude larger than the one of $\mathrm{HIS_{102}}$, whereas \textsc{Propka} predicts both values to be below 6.0 $\mathrm{pK_a}$\xspace units, and \textsc{Pypka} predicts both values to be greater than 6.0, respectively. 
Furthermore, the $\mathrm{pK_a}$\xspace of $\mathrm{ASP_{8}}$ in Barnase is approximately one order of magnitude decreased compared to the isolated amino acid, a property that is accurately predicted by ASAP. 
However, the predicted uncertainty for $\mathrm{GLU_{73}}$ is comparatively high, owing to the fact that the considerably low $\mathrm{pK_a}$\xspace value of this site exceeds the range covered by our suggested training set, which can, however, be further extended.

\section{Conclusions}
Molecular simulation of chemical reactions in nanoscopic environments requires accurate, atomistically resolved molecular structures, because molecular function is tightly coupled to the electronic structure to which every individual atom and electron contribute decisively. However, experimentally determined structures are often subject to significant uncertainties. Hence, structures obtained by experiment must be carefully analyzed and processed before being considered as a starting point for computational modeling. However, this cannot be done manually for routine and high-throughput applications. Our ASAP workflow for structure analysis and protonation achieves automation to a high degree and has been optimized for biomolecular environments. It generates an accurate structural model from
experimental structure information in a fast and efficient manner. Moreover, it combines sequence analysis, error detection, machine learning protonation and p$K_\mathrm{a}$\xspace prediction as well as solvation in a single 
framework. The ASAP workflow seamlessly integrates with our automated SFAM molecular mechanics and hybrid model construction approach, where the corresponding system-focused force field and a quantum region can be generated for the ASAP-prepared structure, which offers a physically consistent protocol for QM/MM modeling. 
        
\section*{Acknowledgements}
The authors gratefully acknowledge financial support through SNF project no. 200021\_182400.
Moreover, K.-S.C. was supported by ETH Zurich and SNF grant no. 200021\_172950-1 awarded to PD Dr. Thomas Hofstetter.

\section*{Supporting Information}
PDB identifiers of the ASAP test set, Cartesian coordinates of the training set for p$K_\mathrm{a}$\xspace prediction, and additional information about the calculation of p$K_\mathrm{a}$\xspace values with semiempirical electronic structure methods, and without implicit solvent models, are provided as additional material in the Supporting Information.

\section*{Data Availability Statement}
ASAP has been implemented in our open-source software \textsc{Scine Swoose}.\cite{swoose1.0.0}
\textsc{Scine Utilities} and \textsc{Scine Molassembler} are available in Refs. \citenum{Utils2023} and \citenum{Molassembler2023}. The \textsc{Scine} framework is also
available from GitHub (\url{https://github.com/qcscine}). Additionally, all Cartesian coordinates of training and test set molecules, and data generated by ASAP can be found in the \textsc{ASAPDATA} repository on Zenodo.


\begin{mcitethebibliography}{101}
\providecommand*\natexlab[1]{#1}
\providecommand*\mciteSetBstSublistMode[1]{}
\providecommand*\mciteSetBstMaxWidthForm[2]{}
\providecommand*\mciteBstWouldAddEndPuncttrue
  {\def\EndOfBibitem{\unskip.}}
\providecommand*\mciteBstWouldAddEndPunctfalse
  {\let\EndOfBibitem\relax}
\providecommand*\mciteSetBstMidEndSepPunct[3]{}
\providecommand*\mciteSetBstSublistLabelBeginEnd[3]{}
\providecommand*\EndOfBibitem{}
\mciteSetBstSublistMode{f}
\mciteSetBstMaxWidthForm{subitem}{(\alph{mcitesubitemcount})}
\mciteSetBstSublistLabelBeginEnd
  {\mcitemaxwidthsubitemform\space}
  {\relax}
  {\relax}

\bibitem[Wodak \latin{et~al.}(2019)Wodak, Paci, Dokholyan, Berezovsky,
  Horovitz, Li, Hilser, Bahar, Karanicolas, Stock, Hamm, Stote, Eberhardt,
  Chebaro, Dejaegere, Cecchini, Changeux, Bolhuis, Vreede, Faccioli, Orioli,
  Ravasio, Yan, Brito, Wyart, Gkeka, Rivalta, Palermo, McCammon,
  Panecka-Hofman, Wade, {Di Pizio}, Niv, Nussinov, Tsai, Jang, Padhorny,
  Kozakov, and McLeish]{Wodak2019}
Wodak,~S.~J.; Paci,~E.; Dokholyan,~N.~V.; Berezovsky,~I.~N.; Horovitz,~A.;
  Li,~J.; Hilser,~V.~J.; Bahar,~I.; Karanicolas,~J.; Stock,~G.; Hamm,~P.;
  Stote,~R.~H.; Eberhardt,~J.; Chebaro,~Y.; Dejaegere,~A.; Cecchini,~M.;
  Changeux,~J.-P.; Bolhuis,~P.~G.; Vreede,~J.; Faccioli,~P.; Orioli,~S.;
  Ravasio,~R.; Yan,~L.; Brito,~C.; Wyart,~M.; Gkeka,~P.; Rivalta,~I.;
  Palermo,~G.; McCammon,~J.~A.; Panecka-Hofman,~J.; Wade,~R.~C.; {Di
  Pizio},~A.; Niv,~M.~Y.; Nussinov,~R.; Tsai,~C.-J.; Jang,~H.; Padhorny,~D.;
  Kozakov,~D.; McLeish,~T. Allostery in Its Many Disguises: From Theory to
  Applications. \emph{Structure} \textbf{2019}, \emph{27}, 566--578\relax
\mciteBstWouldAddEndPuncttrue
\mciteSetBstMidEndSepPunct{\mcitedefaultmidpunct}
{\mcitedefaultendpunct}{\mcitedefaultseppunct}\relax
\EndOfBibitem
\bibitem[Zhou \latin{et~al.}(2019)Zhou, Musalgaonkar, Johnson, and
  Taylor]{Zhou2019}
Zhou,~Y.; Musalgaonkar,~S.; Johnson,~A.~W.; Taylor,~D.~W. {Tightly-orchestrated
  rearrangements govern catalytic center assembly of the ribosome}. \emph{Nat.
  Commun.} \textbf{2019}, \emph{10}, 1--11\relax
\mciteBstWouldAddEndPuncttrue
\mciteSetBstMidEndSepPunct{\mcitedefaultmidpunct}
{\mcitedefaultendpunct}{\mcitedefaultseppunct}\relax
\EndOfBibitem
\bibitem[Garcia-Viloca \latin{et~al.}(2004)Garcia-Viloca, Poulsen, Truhlar, and
  Gao]{Garcia-Viloca2004}
Garcia-Viloca,~M.; Poulsen,~T.~D.; Truhlar,~D.~G.; Gao,~J. {Sensitivity of
  molecular dynamics simulations to the choice of the X-ray structure used to
  model an enzymatic reaction}. \emph{Protein Sci.} \textbf{2004}, \emph{13},
  2341--2354\relax
\mciteBstWouldAddEndPuncttrue
\mciteSetBstMidEndSepPunct{\mcitedefaultmidpunct}
{\mcitedefaultendpunct}{\mcitedefaultseppunct}\relax
\EndOfBibitem
\bibitem[Senior \latin{et~al.}(2020)Senior, Evans, Jumper, Kirkpatrick, Sifre,
  Green, Qin, {\v{Z}}{\'{i}}dek, Nelson, Bridgland, Penedones, Petersen,
  Simonyan, Crossan, Kohli, Jones, Silver, Kavukcuoglu, and
  Hassabis]{Senior2020}
Senior,~A.~W.; Evans,~R.; Jumper,~J.; Kirkpatrick,~J.; Sifre,~L.; Green,~T.;
  Qin,~C.; {\v{Z}}{\'{i}}dek,~A.; Nelson,~A.~W.; Bridgland,~A.; Penedones,~H.;
  Petersen,~S.; Simonyan,~K.; Crossan,~S.; Kohli,~P.; Jones,~D.~T.; Silver,~D.;
  Kavukcuoglu,~K.; Hassabis,~D. {Improved protein structure prediction using
  potentials from deep learning}. \emph{Nature} \textbf{2020}, \emph{577},
  706--710\relax
\mciteBstWouldAddEndPuncttrue
\mciteSetBstMidEndSepPunct{\mcitedefaultmidpunct}
{\mcitedefaultendpunct}{\mcitedefaultseppunct}\relax
\EndOfBibitem
\bibitem[Jumper \latin{et~al.}(2021)Jumper, Evans, Pritzel, Green, Figurnov,
  Ronneberger, Tunyasuvunakool, Bates, {\v{Z}}{\'{i}}dek, Potapenko, Bridgland,
  Meyer, Kohl, Ballard, Cowie, Romera-Paredes, Nikolov, Jain, Adler, Back,
  Petersen, Reiman, Clancy, Zielinski, Steinegger, Pacholska, Berghammer,
  Bodenstein, Silver, Vinyals, Senior, Kavukcuoglu, Kohli, and
  Hassabis]{Jumper2021}
Jumper,~J.; Evans,~R.; Pritzel,~A.; Green,~T.; Figurnov,~M.; Ronneberger,~O.;
  Tunyasuvunakool,~K.; Bates,~R.; {\v{Z}}{\'{i}}dek,~A.; Potapenko,~A.;
  Bridgland,~A.; Meyer,~C.; Kohl,~S.~A.; Ballard,~A.~J.; Cowie,~A.;
  Romera-Paredes,~B.; Nikolov,~S.; Jain,~R.; Adler,~J.; Back,~T.; Petersen,~S.;
  Reiman,~D.; Clancy,~E.; Zielinski,~M.; Steinegger,~M.; Pacholska,~M.;
  Berghammer,~T.; Bodenstein,~S.; Silver,~D.; Vinyals,~O.; Senior,~A.~W.;
  Kavukcuoglu,~K.; Kohli,~P.; Hassabis,~D. {Highly accurate protein structure
  prediction with AlphaFold}. \emph{Nature} \textbf{2021}, \emph{596},
  583--589\relax
\mciteBstWouldAddEndPuncttrue
\mciteSetBstMidEndSepPunct{\mcitedefaultmidpunct}
{\mcitedefaultendpunct}{\mcitedefaultseppunct}\relax
\EndOfBibitem
\bibitem[Baek \latin{et~al.}(2021)Baek, DiMaio, Anishchenko, Dauparas,
  Ovchinnikov, Lee, Wang, Cong, Kinch, Schaeffer, Mill{\'{a}}n, Park, Adams,
  Glassman, DeGiovanni, Pereira, Rodrigues, van Dijk, Ebrecht, Opperman,
  Sagmeister, Buhlheller, Pavkov-Keller, Rathinaswamy, Dalwadi, Yip, Burke,
  Garcia, Grishin, Adams, Read, and Baker]{Baek2021}
Baek,~M.; DiMaio,~F.; Anishchenko,~I.; Dauparas,~J.; Ovchinnikov,~S.;
  Lee,~G.~R.; Wang,~J.; Cong,~Q.; Kinch,~L.~N.; Schaeffer,~R.~D.;
  Mill{\'{a}}n,~C.; Park,~H.; Adams,~C.; Glassman,~C.~R.; DeGiovanni,~A.;
  Pereira,~J.~H.; Rodrigues,~A.~V.; van Dijk,~A.~A.; Ebrecht,~A.~C.;
  Opperman,~D.~J.; Sagmeister,~T.; Buhlheller,~C.; Pavkov-Keller,~T.;
  Rathinaswamy,~M.~K.; Dalwadi,~U.; Yip,~C.~K.; Burke,~J.~E.; Garcia,~K.~C.;
  Grishin,~N.~V.; Adams,~P.~D.; Read,~R.~J.; Baker,~D. {Accurate prediction of
  protein structures and interactions using a three-track neural network}.
  \emph{Science} \textbf{2021}, \emph{373}, 871--876\relax
\mciteBstWouldAddEndPuncttrue
\mciteSetBstMidEndSepPunct{\mcitedefaultmidpunct}
{\mcitedefaultendpunct}{\mcitedefaultseppunct}\relax
\EndOfBibitem
\bibitem[Berman(2000)]{Berman2000}
Berman,~H.~M. {The Protein Data Bank}. \emph{Nucleic Acids Res.} \textbf{2000},
  \emph{28}, 235--242\relax
\mciteBstWouldAddEndPuncttrue
\mciteSetBstMidEndSepPunct{\mcitedefaultmidpunct}
{\mcitedefaultendpunct}{\mcitedefaultseppunct}\relax
\EndOfBibitem
\bibitem[Longenecker \latin{et~al.}(2001)Longenecker, Garrard, Sheffield, and
  Derewenda]{Longenecker2001}
Longenecker,~K.~L.; Garrard,~S.~M.; Sheffield,~P.~J.; Derewenda,~Z.~S. {Protein
  crystallization by rational mutagenesis of surface residues: Lys to Ala
  mutations promote crystallization of RhoGDI}. \emph{Acta Crystallogr. Sect. D
  Biol. Crystallogr.} \textbf{2001}, \emph{57}, 679--688\relax
\mciteBstWouldAddEndPuncttrue
\mciteSetBstMidEndSepPunct{\mcitedefaultmidpunct}
{\mcitedefaultendpunct}{\mcitedefaultseppunct}\relax
\EndOfBibitem
\bibitem[Brunken \latin{et~al.}(2021)Brunken, Csizi, and Reiher]{swoose1.0.0}
Brunken,~C.; Csizi,~K.-S.; Reiher,~M. {qcscine/swoose: Release 1.0.0}.
  2021\relax
\mciteBstWouldAddEndPuncttrue
\mciteSetBstMidEndSepPunct{\mcitedefaultmidpunct}
{\mcitedefaultendpunct}{\mcitedefaultseppunct}\relax
\EndOfBibitem
\bibitem[Reiher \latin{et~al.}()Reiher, \latin{et~al.} others]{scine}
Reiher,~M., \latin{et~al.}  {SCINE $-$ Software for Chemical Interaction
  Networks}. \url{http://scine.ethz.ch/}, visited on 2021-10-26\relax
\mciteBstWouldAddEndPuncttrue
\mciteSetBstMidEndSepPunct{\mcitedefaultmidpunct}
{\mcitedefaultendpunct}{\mcitedefaultseppunct}\relax
\EndOfBibitem
\bibitem[{Madhavi Sastry} \latin{et~al.}(2013){Madhavi Sastry}, Adzhigirey,
  Day, Annabhimoju, and Sherman]{Sastry2013}
{Madhavi Sastry},~G.; Adzhigirey,~M.; Day,~T.; Annabhimoju,~R.; Sherman,~W.
  {Protein and ligand preparation: parameters, protocols, and influence on
  virtual screening enrichments}. \emph{J. Comput. Aided. Mol. Des.}
  \textbf{2013}, \emph{27}, 221--234\relax
\mciteBstWouldAddEndPuncttrue
\mciteSetBstMidEndSepPunct{\mcitedefaultmidpunct}
{\mcitedefaultendpunct}{\mcitedefaultseppunct}\relax
\EndOfBibitem
\bibitem[Eastman \latin{et~al.}(2017)Eastman, Swails, Chodera, McGibbon, Zhao,
  Beauchamp, Wang, Simmonett, Harrigan, Stern, Wiewiora, Brooks, and
  Pande]{openMM}
Eastman,~P.; Swails,~J.; Chodera,~J.~D.; McGibbon,~R.~T.; Zhao,~Y.;
  Beauchamp,~K.~A.; Wang,~L.-P.; Simmonett,~A.~C.; Harrigan,~M.~P.;
  Stern,~C.~D.; Wiewiora,~R.~P.; Brooks,~B.~R.; Pande,~V.~S. OpenMM 7: Rapid
  development of high performance algorithms for molecular dynamics. \emph{PLOS
  Computational Biology} \textbf{2017}, \emph{13}, 1--17\relax
\mciteBstWouldAddEndPuncttrue
\mciteSetBstMidEndSepPunct{\mcitedefaultmidpunct}
{\mcitedefaultendpunct}{\mcitedefaultseppunct}\relax
\EndOfBibitem
\bibitem[Proppe \latin{et~al.}(2016)Proppe, Husch, Simm, and
  Reiher]{Proppe2016}
Proppe,~J.; Husch,~T.; Simm,~G.~N.; Reiher,~M. {Uncertainty quantification for
  quantum chemical models of complex reaction networks}. \emph{Faraday
  Discuss.} \textbf{2016}, \emph{195}, 497--520\relax
\mciteBstWouldAddEndPuncttrue
\mciteSetBstMidEndSepPunct{\mcitedefaultmidpunct}
{\mcitedefaultendpunct}{\mcitedefaultseppunct}\relax
\EndOfBibitem
\bibitem[Simm \latin{et~al.}(2017)Simm, Proppe, and Reiher]{Simm2017}
Simm,~G.~N.; Proppe,~J.; Reiher,~M. {Error Assessment of Computational Models
  in Chemistry}. \emph{Chimia} \textbf{2017}, \emph{71}, 202\relax
\mciteBstWouldAddEndPuncttrue
\mciteSetBstMidEndSepPunct{\mcitedefaultmidpunct}
{\mcitedefaultendpunct}{\mcitedefaultseppunct}\relax
\EndOfBibitem
\bibitem[Brunken and Reiher(2021)Brunken, and Reiher]{Brunken2021}
Brunken,~C.; Reiher,~M. {Automated Construction of Quantum–Classical Hybrid
  Models}. \emph{J. Chem. Theory Comput.} \textbf{2021}, \emph{17},
  3797--3813\relax
\mciteBstWouldAddEndPuncttrue
\mciteSetBstMidEndSepPunct{\mcitedefaultmidpunct}
{\mcitedefaultendpunct}{\mcitedefaultseppunct}\relax
\EndOfBibitem
\bibitem[Brunken and Reiher(2020)Brunken, and Reiher]{Brunken2020}
Brunken,~C.; Reiher,~M. {Self-Parametrizing System-Focused Atomistic Models}.
  \emph{J. Chem. Theory Comput.} \textbf{2020}, \emph{16}, 1646--1665\relax
\mciteBstWouldAddEndPuncttrue
\mciteSetBstMidEndSepPunct{\mcitedefaultmidpunct}
{\mcitedefaultendpunct}{\mcitedefaultseppunct}\relax
\EndOfBibitem
\bibitem[Alongi and Shields(2010)Alongi, and Shields]{Alongi2010}
Alongi,~K.~S.; Shields,~G.~C. \emph{Annual Reports in Computational Chemistry};
  Elsevier: Amsterdam, The Netherlands, 2010; pp 113--138\relax
\mciteBstWouldAddEndPuncttrue
\mciteSetBstMidEndSepPunct{\mcitedefaultmidpunct}
{\mcitedefaultendpunct}{\mcitedefaultseppunct}\relax
\EndOfBibitem
\bibitem[Ho and Coote(2009)Ho, and Coote]{Ho2009}
Ho,~J.; Coote,~M.~L. {A universal approach for continuum solvent pKa
  calculations: Are we there yet?} \emph{Theor. Chem. Acc.} \textbf{2009},
  \emph{125}, 3--21\relax
\mciteBstWouldAddEndPuncttrue
\mciteSetBstMidEndSepPunct{\mcitedefaultmidpunct}
{\mcitedefaultendpunct}{\mcitedefaultseppunct}\relax
\EndOfBibitem
\bibitem[Casasnovas \latin{et~al.}(2014)Casasnovas, Ortega-Castro, Frau,
  Donoso, and Mu{\~{n}}oz]{Casasnovas2014}
Casasnovas,~R.; Ortega-Castro,~J.; Frau,~J.; Donoso,~J.; Mu{\~{n}}oz,~F.
  {Theoretical pKa calculations with continuum model solvents, alternative
  protocols to thermodynamic cycles}. \textbf{2014}, \emph{114},
  1350--1363\relax
\mciteBstWouldAddEndPuncttrue
\mciteSetBstMidEndSepPunct{\mcitedefaultmidpunct}
{\mcitedefaultendpunct}{\mcitedefaultseppunct}\relax
\EndOfBibitem
\bibitem[Ho(2015)]{Ho2015}
Ho,~J. {Are thermodynamic cycles necessary for continuum solvent calculation of
  pKas and reduction potentials?} \emph{Phys. Chem. Chem. Phys.} \textbf{2015},
  \emph{17}, 2859--2868\relax
\mciteBstWouldAddEndPuncttrue
\mciteSetBstMidEndSepPunct{\mcitedefaultmidpunct}
{\mcitedefaultendpunct}{\mcitedefaultseppunct}\relax
\EndOfBibitem
\bibitem[Seybold and Shields(2015)Seybold, and Shields]{Seybold2015}
Seybold,~P.~G.; Shields,~G.~C. {Computational estimation of pKa values}.
  \emph{Wiley Interdiscip. Rev. Comput. Mol. Sci.} \textbf{2015}, \emph{5},
  290--297\relax
\mciteBstWouldAddEndPuncttrue
\mciteSetBstMidEndSepPunct{\mcitedefaultmidpunct}
{\mcitedefaultendpunct}{\mcitedefaultseppunct}\relax
\EndOfBibitem
\bibitem[Bashford and Karplus(1990)Bashford, and Karplus]{Bashford1990}
Bashford,~D.; Karplus,~M. {pKa's of Ionizable Groups in Proteins: Atomic Detail
  from a Continuum Electrostatic Model}. \emph{Biochemistry} \textbf{1990},
  \emph{29}, 10219--10225\relax
\mciteBstWouldAddEndPuncttrue
\mciteSetBstMidEndSepPunct{\mcitedefaultmidpunct}
{\mcitedefaultendpunct}{\mcitedefaultseppunct}\relax
\EndOfBibitem
\bibitem[Wang \latin{et~al.}(2015)Wang, Li, and Alexov]{Wang2015a}
Wang,~L.; Li,~L.; Alexov,~E. {pKa predictions for proteins, RNAs, and DNAs with
  the Gaussian dielectric function using DelPhi pKa}. \emph{Proteins Struct.
  Funct. Bioinforma.} \textbf{2015}, \emph{83}, 2186--2197\relax
\mciteBstWouldAddEndPuncttrue
\mciteSetBstMidEndSepPunct{\mcitedefaultmidpunct}
{\mcitedefaultendpunct}{\mcitedefaultseppunct}\relax
\EndOfBibitem
\bibitem[Ho(2014)]{Ho2014}
Ho,~J. {Predicting pKa in implicit solvents: Current status and future
  directions}. \emph{{Aust. J. Chem.}} \textbf{2014}, \emph{67},
  1441--1460\relax
\mciteBstWouldAddEndPuncttrue
\mciteSetBstMidEndSepPunct{\mcitedefaultmidpunct}
{\mcitedefaultendpunct}{\mcitedefaultseppunct}\relax
\EndOfBibitem
\bibitem[Klici{\'{c}} \latin{et~al.}(2002)Klici{\'{c}}, Friesner, Liu, and
  Guida]{Klicic2002}
Klici{\'{c}},~J.~J.; Friesner,~R.~A.; Liu,~S.~Y.; Guida,~W.~C. {Accurate
  prediction of acidity constants in aqueous solution via density functional
  theory and self-consistent reaction field methods}. \emph{J. Phys. Chem. A}
  \textbf{2002}, \emph{106}, 1327--1335\relax
\mciteBstWouldAddEndPuncttrue
\mciteSetBstMidEndSepPunct{\mcitedefaultmidpunct}
{\mcitedefaultendpunct}{\mcitedefaultseppunct}\relax
\EndOfBibitem
\bibitem[Klamt \latin{et~al.}(2003)Klamt, Eckert, Diedenhofen, and
  Beck]{Klamt2003}
Klamt,~A.; Eckert,~F.; Diedenhofen,~M.; Beck,~M.~E. {First Principles
  Calculations of Aqueous pKa Values for Organic and Inorganic Acids Using
  COSMO-RS Reveal an Inconsistency in the Slope of the pKa Scale}. \emph{J.
  Phys. Chem. A} \textbf{2003}, \emph{107}, 9380--9386\relax
\mciteBstWouldAddEndPuncttrue
\mciteSetBstMidEndSepPunct{\mcitedefaultmidpunct}
{\mcitedefaultendpunct}{\mcitedefaultseppunct}\relax
\EndOfBibitem
\bibitem[Riojas and Wilson(2014)Riojas, and Wilson]{Riojas2014}
Riojas,~A.~G.; Wilson,~A.~K. {Solv-ccCA: Implicit solvation and the correlation
  consistent composite approach for the determination of p Ka}. \emph{J. Chem.
  Theory Comput.} \textbf{2014}, \emph{10}, 1500--1510\relax
\mciteBstWouldAddEndPuncttrue
\mciteSetBstMidEndSepPunct{\mcitedefaultmidpunct}
{\mcitedefaultendpunct}{\mcitedefaultseppunct}\relax
\EndOfBibitem
\bibitem[Thapa and Schlegel(2017)Thapa, and Schlegel]{Thapa2017}
Thapa,~B.; Schlegel,~H.~B. {Improved pKa Prediction of Substituted Alcohols,
  Phenols, and Hydroperoxides in Aqueous Medium Using Density Functional Theory
  and a Cluster-Continuum Solvation Model}. \emph{J. Phys. Chem. A}
  \textbf{2017}, \emph{121}, 4698--4706\relax
\mciteBstWouldAddEndPuncttrue
\mciteSetBstMidEndSepPunct{\mcitedefaultmidpunct}
{\mcitedefaultendpunct}{\mcitedefaultseppunct}\relax
\EndOfBibitem
\bibitem[Lian \latin{et~al.}(2018)Lian, Johnston, Parks, and Smith]{Lian2018}
Lian,~P.; Johnston,~R.~C.; Parks,~J.~M.; Smith,~J.~C. {Quantum Chemical
  Calculation of p Kas of Environmentally Relevant Functional Groups:
  Carboxylic Acids, Amines, and Thiols in Aqueous Solution}. \emph{J. Phys.
  Chem. A} \textbf{2018}, \emph{122}, 4366--4374\relax
\mciteBstWouldAddEndPuncttrue
\mciteSetBstMidEndSepPunct{\mcitedefaultmidpunct}
{\mcitedefaultendpunct}{\mcitedefaultseppunct}\relax
\EndOfBibitem
\bibitem[Donnini \latin{et~al.}(2011)Donnini, Tegeler, Groenhof, and
  Grubm{\"{u}}ller]{Donnini2011}
Donnini,~S.; Tegeler,~F.; Groenhof,~G.; Grubm{\"{u}}ller,~H. {Constant pH
  Molecular Dynamics in Explicit Solvent with $\lambda$-Dynamics}. \emph{J.
  Chem. Theory Comput.} \textbf{2011}, \emph{7}, 1962--1978\relax
\mciteBstWouldAddEndPuncttrue
\mciteSetBstMidEndSepPunct{\mcitedefaultmidpunct}
{\mcitedefaultendpunct}{\mcitedefaultseppunct}\relax
\EndOfBibitem
\bibitem[Itoh \latin{et~al.}(2011)Itoh, Damjanovi{\'{c}}, and Brooks]{Itoh2011}
Itoh,~S.~G.; Damjanovi{\'{c}},~A.; Brooks,~B.~R. {pH replica-exchange method
  based on discrete protonation states}. \emph{Proteins Struct. Funct.
  Bioinforma.} \textbf{2011}, \emph{79}, 3420--3436\relax
\mciteBstWouldAddEndPuncttrue
\mciteSetBstMidEndSepPunct{\mcitedefaultmidpunct}
{\mcitedefaultendpunct}{\mcitedefaultseppunct}\relax
\EndOfBibitem
\bibitem[Wallace and Shen(2011)Wallace, and Shen]{Wallace2011}
Wallace,~J.~A.; Shen,~J.~K. {Continuous Constant pH Molecular Dynamics in
  Explicit Solvent with pH-Based Replica Exchange}. \emph{J. Chem. Theory
  Comput.} \textbf{2011}, \emph{7}, 2617--2629\relax
\mciteBstWouldAddEndPuncttrue
\mciteSetBstMidEndSepPunct{\mcitedefaultmidpunct}
{\mcitedefaultendpunct}{\mcitedefaultseppunct}\relax
\EndOfBibitem
\bibitem[Huang \latin{et~al.}(2016)Huang, Chen, Wallace, and Shen]{Huang2016a}
Huang,~Y.; Chen,~W.; Wallace,~J.~A.; Shen,~J. {All-Atom Continuous Constant pH
  Molecular Dynamics With Particle Mesh Ewald and Titratable Water}. \emph{J.
  Chem. Theory Comput.} \textbf{2016}, \emph{12}, 5411--5421\relax
\mciteBstWouldAddEndPuncttrue
\mciteSetBstMidEndSepPunct{\mcitedefaultmidpunct}
{\mcitedefaultendpunct}{\mcitedefaultseppunct}\relax
\EndOfBibitem
\bibitem[Li \latin{et~al.}(2002)Li, Hains, Everts, Robertson, and
  Jensen]{Li2002}
Li,~H.; Hains,~A.~W.; Everts,~J.~E.; Robertson,~A.~D.; Jensen,~J.~H. {The
  Prediction of Protein p K a 's Using QM/MM: The p K a of Lysine 55 in Turkey
  Ovomucoid Third Domain}. \emph{J. Phys. Chem. B} \textbf{2002}, \emph{106},
  3486--3494\relax
\mciteBstWouldAddEndPuncttrue
\mciteSetBstMidEndSepPunct{\mcitedefaultmidpunct}
{\mcitedefaultendpunct}{\mcitedefaultseppunct}\relax
\EndOfBibitem
\bibitem[Li and Cui(2003)Li, and Cui]{Li2003}
Li,~G.; Cui,~Q. {p K a Calculations with QM/MM Free Energy Perturbations}.
  \emph{J. Phys. Chem. B} \textbf{2003}, \emph{107}, 14521--14528\relax
\mciteBstWouldAddEndPuncttrue
\mciteSetBstMidEndSepPunct{\mcitedefaultmidpunct}
{\mcitedefaultendpunct}{\mcitedefaultseppunct}\relax
\EndOfBibitem
\bibitem[Ho and Coote(2011)Ho, and Coote]{Ho2011}
Ho,~J.; Coote,~M.~L. {First‐principles prediction of acidities in the gas and
  solution phase}. \emph{WIREs Comput. Mol. Sci.} \textbf{2011}, \emph{1},
  649--660\relax
\mciteBstWouldAddEndPuncttrue
\mciteSetBstMidEndSepPunct{\mcitedefaultmidpunct}
{\mcitedefaultendpunct}{\mcitedefaultseppunct}\relax
\EndOfBibitem
\bibitem[Godoy-Ruiz \latin{et~al.}(2005)Godoy-Ruiz, Perez-Jimenez, Garcia-Mira,
  {Plaza del Pino}, and Sanchez-Ruiz]{Godoy-Ruiz2005}
Godoy-Ruiz,~R.; Perez-Jimenez,~R.; Garcia-Mira,~M.~M.; {Plaza del Pino},~I.~M.;
  Sanchez-Ruiz,~J.~M. Empirical parametrization of pK values for carboxylic
  acids in proteins using a genetic algorithm. \emph{Biophysical Chemistry}
  \textbf{2005}, \emph{115}, 263--266\relax
\mciteBstWouldAddEndPuncttrue
\mciteSetBstMidEndSepPunct{\mcitedefaultmidpunct}
{\mcitedefaultendpunct}{\mcitedefaultseppunct}\relax
\EndOfBibitem
\bibitem[Li \latin{et~al.}(2005)Li, Robertson, and Jensen]{Li2005}
Li,~H.; Robertson,~A.~D.; Jensen,~J.~H. {Very fast empirical prediction and
  rationalization of protein pK a values}. \emph{Proteins Struct. Funct.
  Genet.} \textbf{2005}, \emph{61}, 704--721\relax
\mciteBstWouldAddEndPuncttrue
\mciteSetBstMidEndSepPunct{\mcitedefaultmidpunct}
{\mcitedefaultendpunct}{\mcitedefaultseppunct}\relax
\EndOfBibitem
\bibitem[Krieger \latin{et~al.}(2006)Krieger, Nielsen, Spronk, and
  Vriend]{Krieger2006}
Krieger,~E.; Nielsen,~J.~E.; Spronk,~C.~A.; Vriend,~G. {Fast empirical pKa
  prediction by Ewald summation}. \emph{J. Mol. Graph. Model.} \textbf{2006},
  \emph{25}, 481--486\relax
\mciteBstWouldAddEndPuncttrue
\mciteSetBstMidEndSepPunct{\mcitedefaultmidpunct}
{\mcitedefaultendpunct}{\mcitedefaultseppunct}\relax
\EndOfBibitem
\bibitem[Bas \latin{et~al.}(2008)Bas, Rogers, and Jensen]{Bas2008b}
Bas,~D.~C.; Rogers,~D.~M.; Jensen,~J.~H. {Very fast prediction and
  rationalization of pKa values for protein-ligand complexes}. \emph{Proteins
  Struct. Funct. Genet.} \textbf{2008}, \emph{73}, 765--783\relax
\mciteBstWouldAddEndPuncttrue
\mciteSetBstMidEndSepPunct{\mcitedefaultmidpunct}
{\mcitedefaultendpunct}{\mcitedefaultseppunct}\relax
\EndOfBibitem
\bibitem[Reis \latin{et~al.}(2020)Reis, Vila-Vicosa, Rocchia, and
  Machuqueiro]{Reis2020}
Reis,~P.~B.; Vila-Vicosa,~D.; Rocchia,~W.; Machuqueiro,~M. {PypKA: A flexible
  python module for poisson-Boltzmann-based pKa calculations}. \emph{J. Chem.
  Inf. Model.} \textbf{2020}, \emph{60}, 4442--4448\relax
\mciteBstWouldAddEndPuncttrue
\mciteSetBstMidEndSepPunct{\mcitedefaultmidpunct}
{\mcitedefaultendpunct}{\mcitedefaultseppunct}\relax
\EndOfBibitem
\bibitem[Dolinsky \latin{et~al.}(2004)Dolinsky, Nielsen, McCammon, and
  Baker]{Dolinsky2004}
Dolinsky,~T.~J.; Nielsen,~J.~E.; McCammon,~J.~A.; Baker,~N.~A. {PDB2PQR: An
  automated pipeline for the setup of Poisson-Boltzmann electrostatics
  calculations}. \emph{Nucleic Acids Res.} \textbf{2004}, \emph{32}\relax
\mciteBstWouldAddEndPuncttrue
\mciteSetBstMidEndSepPunct{\mcitedefaultmidpunct}
{\mcitedefaultendpunct}{\mcitedefaultseppunct}\relax
\EndOfBibitem
\bibitem[Dolinsky \latin{et~al.}(2007)Dolinsky, Czodrowski, Li, Nielsen,
  Jensen, Klebe, and Baker]{Dolinsky2007}
Dolinsky,~T.~J.; Czodrowski,~P.; Li,~H.; Nielsen,~J.~E.; Jensen,~J.~H.;
  Klebe,~G.; Baker,~N.~A. {PDB2PQR: Expanding and upgrading automated
  preparation of biomolecular structures for molecular simulations}.
  \emph{Nucleic Acids Res.} \textbf{2007}, \emph{35}, W522\relax
\mciteBstWouldAddEndPuncttrue
\mciteSetBstMidEndSepPunct{\mcitedefaultmidpunct}
{\mcitedefaultendpunct}{\mcitedefaultseppunct}\relax
\EndOfBibitem
\bibitem[Anandakrishnan \latin{et~al.}(2012)Anandakrishnan, Aguilar, and
  Onufriev]{Anandakrishnan2012}
Anandakrishnan,~R.; Aguilar,~B.; Onufriev,~A.~V. {H++ 3.0: Automating pK
  prediction and the preparation of biomolecular structures for atomistic
  molecular modeling and simulations}. \emph{Nucleic Acids Res.} \textbf{2012},
  \emph{40}, 537--541\relax
\mciteBstWouldAddEndPuncttrue
\mciteSetBstMidEndSepPunct{\mcitedefaultmidpunct}
{\mcitedefaultendpunct}{\mcitedefaultseppunct}\relax
\EndOfBibitem
\bibitem[Olsson \latin{et~al.}(2011)Olsson, S{\o}ndergaard, Rostkowski, and
  Jensen]{Olsson2011}
Olsson,~M.~H.; S{\o}ndergaard,~C.~R.; Rostkowski,~M.; Jensen,~J.~H. {PROPKA3:
  Consistent treatment of internal and surface residues in empirical p K a
  predictions}. \emph{J. Chem. Theory Comput.} \textbf{2011}, \emph{7},
  525--537\relax
\mciteBstWouldAddEndPuncttrue
\mciteSetBstMidEndSepPunct{\mcitedefaultmidpunct}
{\mcitedefaultendpunct}{\mcitedefaultseppunct}\relax
\EndOfBibitem
\bibitem[Bochevarov \latin{et~al.}(2016)Bochevarov, Watson, Greenwood, and
  Philipp]{Bochevarov2016a}
Bochevarov,~A.~D.; Watson,~M.~A.; Greenwood,~J.~R.; Philipp,~D.~M.
  {Multiconformation, Density Functional Theory-Based pKa Prediction in
  Application to Large, Flexible Organic Molecules with Diverse Functional
  Groups}. \emph{J. Chem. Theory Comput.} \textbf{2016}, \emph{12},
  6001--6019\relax
\mciteBstWouldAddEndPuncttrue
\mciteSetBstMidEndSepPunct{\mcitedefaultmidpunct}
{\mcitedefaultendpunct}{\mcitedefaultseppunct}\relax
\EndOfBibitem
\bibitem[Ratkova \latin{et~al.}(2015)Ratkova, Palmer, and Fedorov]{Ratkova2015}
Ratkova,~E.~L.; Palmer,~D.~S.; Fedorov,~M.~V. {Solvation Thermodynamics of
  Organic Molecules by the Molecular Integral Equation Theory: Approaching
  Chemical Accuracy}. \emph{Chem. Rev.} \textbf{2015}, \emph{115},
  6312--6356\relax
\mciteBstWouldAddEndPuncttrue
\mciteSetBstMidEndSepPunct{\mcitedefaultmidpunct}
{\mcitedefaultendpunct}{\mcitedefaultseppunct}\relax
\EndOfBibitem
\bibitem[Matsui \latin{et~al.}(2012)Matsui, Baba, Kamiya, and
  Shigeta]{Matsui2012}
Matsui,~T.; Baba,~T.; Kamiya,~K.; Shigeta,~Y. {An accurate density functional
  theory based estimation of pK a values of polar residues combined with
  experimental data: From amino acids to minimal proteins}. \emph{Phys. Chem.
  Chem. Phys.} \textbf{2012}, \emph{14}, 4181--4187\relax
\mciteBstWouldAddEndPuncttrue
\mciteSetBstMidEndSepPunct{\mcitedefaultmidpunct}
{\mcitedefaultendpunct}{\mcitedefaultseppunct}\relax
\EndOfBibitem
\bibitem[Pernot \latin{et~al.}(2015)Pernot, Civalleri, Presti, and
  Savin]{Pernot2015}
Pernot,~P.; Civalleri,~B.; Presti,~D.; Savin,~A. {Prediction Uncertainty of
  Density Functional Approximations for Properties of Crystals with Cubic
  Symmetry}. \emph{J. Phys. Chem. A} \textbf{2015}, \emph{119},
  5288--5304\relax
\mciteBstWouldAddEndPuncttrue
\mciteSetBstMidEndSepPunct{\mcitedefaultmidpunct}
{\mcitedefaultendpunct}{\mcitedefaultseppunct}\relax
\EndOfBibitem
\bibitem[Proppe and Reiher(2017)Proppe, and Reiher]{Proppe2017}
Proppe,~J.; Reiher,~M. {Reliable Estimation of Prediction Uncertainty for
  Physicochemical Property Models}. \emph{J. Chem. Theory Comput.}
  \textbf{2017}, \emph{13}, 3297--3317\relax
\mciteBstWouldAddEndPuncttrue
\mciteSetBstMidEndSepPunct{\mcitedefaultmidpunct}
{\mcitedefaultendpunct}{\mcitedefaultseppunct}\relax
\EndOfBibitem
\bibitem[Simm and Reiher(2016)Simm, and Reiher]{Simm2016}
Simm,~G.~N.; Reiher,~M. {Systematic Error Estimation for Chemical Reaction
  Energies}. \emph{J. Chem. Theory Comput.} \textbf{2016}, \emph{12},
  2762--2773\relax
\mciteBstWouldAddEndPuncttrue
\mciteSetBstMidEndSepPunct{\mcitedefaultmidpunct}
{\mcitedefaultendpunct}{\mcitedefaultseppunct}\relax
\EndOfBibitem
\bibitem[Proppe and Reiher(2019)Proppe, and Reiher]{Proppe2019}
Proppe,~J.; Reiher,~M. {Mechanism Deduction from Noisy Chemical Reaction
  Networks}. \emph{J. Chem. Theory Comput.} \textbf{2019}, \emph{15},
  357--370\relax
\mciteBstWouldAddEndPuncttrue
\mciteSetBstMidEndSepPunct{\mcitedefaultmidpunct}
{\mcitedefaultendpunct}{\mcitedefaultseppunct}\relax
\EndOfBibitem
\bibitem[Reiher(2022)]{Reiher2021}
Reiher,~M. {Molecule‐Specific Uncertainty Quantification in Quantum Chemical
  Studies}. \emph{Isr. J. Chem.} \textbf{2022}, \emph{62}, 1--34\relax
\mciteBstWouldAddEndPuncttrue
\mciteSetBstMidEndSepPunct{\mcitedefaultmidpunct}
{\mcitedefaultendpunct}{\mcitedefaultseppunct}\relax
\EndOfBibitem
\bibitem[Simm and Reiher(2018)Simm, and Reiher]{Simm2018}
Simm,~G.~N.; Reiher,~M. {Error-Controlled Exploration of Chemical Reaction
  Networks with Gaussian Processes}. \emph{J. Chem. Theory Comput.}
  \textbf{2018}, \emph{14}, 5238--5248\relax
\mciteBstWouldAddEndPuncttrue
\mciteSetBstMidEndSepPunct{\mcitedefaultmidpunct}
{\mcitedefaultendpunct}{\mcitedefaultseppunct}\relax
\EndOfBibitem
\bibitem[Proppe \latin{et~al.}(2019)Proppe, Gugler, and Reiher]{Proppe2019a}
Proppe,~J.; Gugler,~S.; Reiher,~M. {Gaussian Process-Based Refinement of
  Dispersion Corrections}. \emph{J. Chem. Theory Comput.} \textbf{2019},
  \emph{15}, 6046--6060\relax
\mciteBstWouldAddEndPuncttrue
\mciteSetBstMidEndSepPunct{\mcitedefaultmidpunct}
{\mcitedefaultendpunct}{\mcitedefaultseppunct}\relax
\EndOfBibitem
\bibitem[Deringer \latin{et~al.}(2021)Deringer, Bart{\'{o}}k, Bernstein,
  Wilkins, Ceriotti, and Cs{\'{a}}nyi]{Deringer2021a}
Deringer,~V.~L.; Bart{\'{o}}k,~A.~P.; Bernstein,~N.; Wilkins,~D.~M.;
  Ceriotti,~M.; Cs{\'{a}}nyi,~G. {Gaussian Process Regression for Materials and
  Molecules}. \emph{{Chem. Rev.}} \textbf{2021}, \emph{121}, 10073--10141\relax
\mciteBstWouldAddEndPuncttrue
\mciteSetBstMidEndSepPunct{\mcitedefaultmidpunct}
{\mcitedefaultendpunct}{\mcitedefaultseppunct}\relax
\EndOfBibitem
\bibitem[Gugler and Reiher(2022)Gugler, and Reiher]{Gugler2022}
Gugler,~S.; Reiher,~M. {Quantum Chemical Roots of Machine-Learning Molecular
  Similarity Descriptors}. \emph{J. Chem. Theory Comput.} \textbf{2022},
  \emph{18}, 6670--6689\relax
\mciteBstWouldAddEndPuncttrue
\mciteSetBstMidEndSepPunct{\mcitedefaultmidpunct}
{\mcitedefaultendpunct}{\mcitedefaultseppunct}\relax
\EndOfBibitem
\bibitem[Neese(2012)]{Neese2012}
Neese,~F. {The ORCA program system}. \emph{Wiley Int. Rev. Comput. Mol. Sci.}
  \textbf{2012}, \emph{2}, 73--78\relax
\mciteBstWouldAddEndPuncttrue
\mciteSetBstMidEndSepPunct{\mcitedefaultmidpunct}
{\mcitedefaultendpunct}{\mcitedefaultseppunct}\relax
\EndOfBibitem
\bibitem[Neese \latin{et~al.}(2020)Neese, Wennmohs, Becker, and
  Riplinger]{Neese2020}
Neese,~F.; Wennmohs,~F.; Becker,~U.; Riplinger,~C. {The ORCA quantum chemistry
  program package}. \emph{J. Chem. Phys.} \textbf{2020}, \emph{152},
  224108\relax
\mciteBstWouldAddEndPuncttrue
\mciteSetBstMidEndSepPunct{\mcitedefaultmidpunct}
{\mcitedefaultendpunct}{\mcitedefaultseppunct}\relax
\EndOfBibitem
\bibitem[Perdew \latin{et~al.}(1996)Perdew, Burke, and Ernzerhof]{Perdew1996a}
Perdew,~J.~P.; Burke,~K.; Ernzerhof,~M. {Generalized gradient approximation
  made simple}. \emph{Phys. Rev. Lett.} \textbf{1996}, \emph{77},
  3865--3868\relax
\mciteBstWouldAddEndPuncttrue
\mciteSetBstMidEndSepPunct{\mcitedefaultmidpunct}
{\mcitedefaultendpunct}{\mcitedefaultseppunct}\relax
\EndOfBibitem
\bibitem[Ernzerhof and Scuseria(1999)Ernzerhof, and Scuseria]{Ernzerhof1999}
Ernzerhof,~M.; Scuseria,~G.~E. {Assessment of the Perdew-Burke-Ernzerhof
  exchange-correlation functional}. \emph{J. Chem. Phys.} \textbf{1999},
  \emph{110}, 5029--5036\relax
\mciteBstWouldAddEndPuncttrue
\mciteSetBstMidEndSepPunct{\mcitedefaultmidpunct}
{\mcitedefaultendpunct}{\mcitedefaultseppunct}\relax
\EndOfBibitem
\bibitem[Adamo and Barone(1999)Adamo, and Barone]{Adamo1999}
Adamo,~C.; Barone,~V. {Toward reliable density functional methods without
  adjustable parameters: The PBE0 model}. \emph{J. Chem. Phys.} \textbf{1999},
  \emph{110}, 6158--6170\relax
\mciteBstWouldAddEndPuncttrue
\mciteSetBstMidEndSepPunct{\mcitedefaultmidpunct}
{\mcitedefaultendpunct}{\mcitedefaultseppunct}\relax
\EndOfBibitem
\bibitem[Sch{\"{a}}fer \latin{et~al.}(1994)Sch{\"{a}}fer, Huber, and
  Ahlrichs]{Schafer1994}
Sch{\"{a}}fer,~A.; Huber,~C.; Ahlrichs,~R. {Fully optimized contracted Gaussian
  basis sets of triple zeta valence quality for atoms Li to Kr}. \emph{J. Chem.
  Phys.} \textbf{1994}, \emph{100}, 5829--5835\relax
\mciteBstWouldAddEndPuncttrue
\mciteSetBstMidEndSepPunct{\mcitedefaultmidpunct}
{\mcitedefaultendpunct}{\mcitedefaultseppunct}\relax
\EndOfBibitem
\bibitem[Eichkorn \latin{et~al.}(1995)Eichkorn, Treutler, {\"{O}}hm,
  H{\"{a}}ser, and Ahlrichs]{Eichkorn1995}
Eichkorn,~K.; Treutler,~O.; {\"{O}}hm,~H.; H{\"{a}}ser,~M.; Ahlrichs,~R.
  {Auxiliary basis sets to approximate Coulomb potentials}. \emph{Chem. Phys.
  Lett.} \textbf{1995}, \emph{240}, 283--290\relax
\mciteBstWouldAddEndPuncttrue
\mciteSetBstMidEndSepPunct{\mcitedefaultmidpunct}
{\mcitedefaultendpunct}{\mcitedefaultseppunct}\relax
\EndOfBibitem
\bibitem[Eichkorn \latin{et~al.}(1997)Eichkorn, Weigend, Treutler, and
  Ahlrichs]{Eichkorn1997}
Eichkorn,~K.; Weigend,~F.; Treutler,~O.; Ahlrichs,~R. {Auxiliary basis sets for
  main row atoms and transition metals and their use to approximate Coulomb
  potentials}. \emph{Theor. Chem. Acc.} \textbf{1997}, \emph{97},
  119--124\relax
\mciteBstWouldAddEndPuncttrue
\mciteSetBstMidEndSepPunct{\mcitedefaultmidpunct}
{\mcitedefaultendpunct}{\mcitedefaultseppunct}\relax
\EndOfBibitem
\bibitem[Grimme \latin{et~al.}(2010)Grimme, Antony, Ehrlich, and
  Krieg]{Grimme2010a}
Grimme,~S.; Antony,~J.; Ehrlich,~S.; Krieg,~H. {A consistent and accurate ab
  initio parametrization of density functional dispersion correction (DFT-D)
  for the 94 elements H-Pu}. \emph{J. Chem. Phys.} \textbf{2010}, \emph{132},
  154104\relax
\mciteBstWouldAddEndPuncttrue
\mciteSetBstMidEndSepPunct{\mcitedefaultmidpunct}
{\mcitedefaultendpunct}{\mcitedefaultseppunct}\relax
\EndOfBibitem
\bibitem[Grimme \latin{et~al.}(2011)Grimme, Ehrlich, and Goerigk]{Grimme2011}
Grimme,~S.; Ehrlich,~S.; Goerigk,~L. {Effect of the damping function in
  dispersion corrected density functional theory}. \emph{J. Comput. Chem.}
  \textbf{2011}, \emph{32}, 1456--1465\relax
\mciteBstWouldAddEndPuncttrue
\mciteSetBstMidEndSepPunct{\mcitedefaultmidpunct}
{\mcitedefaultendpunct}{\mcitedefaultseppunct}\relax
\EndOfBibitem
\bibitem[Marenich \latin{et~al.}(2009)Marenich, Cramer, and
  Truhlar]{Marenich2009}
Marenich,~A.~V.; Cramer,~C.~J.; Truhlar,~D.~G. {Universal Solvation Model Based
  on Solute Electron Density and on a Continuum Model of the Solvent Defined by
  the Bulk Dielectric Constant and Atomic Surface Tensions}. \emph{J. Phys.
  Chem. B} \textbf{2009}, \emph{113}, 6378--6396\relax
\mciteBstWouldAddEndPuncttrue
\mciteSetBstMidEndSepPunct{\mcitedefaultmidpunct}
{\mcitedefaultendpunct}{\mcitedefaultseppunct}\relax
\EndOfBibitem
\bibitem[Sobez and Reiher(2020)Sobez, and Reiher]{Sobez2020}
Sobez,~J.~G.; Reiher,~M. {Molassembler: Molecular Graph Construction,
  Modification, and Conformer Generation for Inorganic and Organic Molecules}.
  \emph{J. Chem. Inf. Model.} \textbf{2020}, \emph{60}, 3884--3900\relax
\mciteBstWouldAddEndPuncttrue
\mciteSetBstMidEndSepPunct{\mcitedefaultmidpunct}
{\mcitedefaultendpunct}{\mcitedefaultseppunct}\relax
\EndOfBibitem
\bibitem[Bensberg \latin{et~al.}(2023)Bensberg, Grimmel, Sobez, Steiner,
  Unsleber, and Reiher]{Molassembler2023}
Bensberg,~M.; Grimmel,~S.~A.; Sobez,~J.-G.; Steiner,~M.; Unsleber,~J.~P.;
  Reiher,~M. {qcscine/Molassembler: Release 2.0.0}. 2023;
  \url{https://zenodo.org/record/7928074}\relax
\mciteBstWouldAddEndPuncttrue
\mciteSetBstMidEndSepPunct{\mcitedefaultmidpunct}
{\mcitedefaultendpunct}{\mcitedefaultseppunct}\relax
\EndOfBibitem
\bibitem[Baiardi \latin{et~al.}(2023)Baiardi, Bensberg, Bosia, Brunken, Csizi,
  Feldmann, Glaser, Grimmel, Gugler, Haag, Heuer, M{\"u}ller, Polonius, Simm,
  Sobez, Steiner, T{\"u}rtscher, Unsleber, Vaucher, Weymuth, and
  Reiher]{Utils2023}
Baiardi,~A.; Bensberg,~M.; Bosia,~F.; Brunken,~C.; Csizi,~K.-S.; Feldmann,~R.;
  Glaser,~N.; Grimmel,~S.~A.; Gugler,~S.; Haag,~M.~P.; Heuer,~M.~A.;
  M{\"u}ller,~C.~H.; Polonius,~S.; Simm,~G.~N.; Sobez,~J.-G.; Steiner,~M.;
  T{\"u}rtscher,~P.~L.; Unsleber,~J.~P.; Vaucher,~A.~C.; Weymuth,~T.;
  Reiher,~M. {qcscine/Utilities: Release 8.0.0}. 2023;
  \url{https://zenodo.org/record/7928050}\relax
\mciteBstWouldAddEndPuncttrue
\mciteSetBstMidEndSepPunct{\mcitedefaultmidpunct}
{\mcitedefaultendpunct}{\mcitedefaultseppunct}\relax
\EndOfBibitem
\bibitem[{The PyMOL Molecular Graphics System, Version 2.0
  Schr{\"{o}}dinger}(2015)]{PyMOL}
{The PyMOL Molecular Graphics System, Version 2.0 Schr{\"{o}}dinger},~L.
  {https://pymol.org (accessed November 14, 2022) -- The PyMOL Molecular
  Graphics System, Version 2.5.2}. 2015\relax
\mciteBstWouldAddEndPuncttrue
\mciteSetBstMidEndSepPunct{\mcitedefaultmidpunct}
{\mcitedefaultendpunct}{\mcitedefaultseppunct}\relax
\EndOfBibitem
\bibitem[Bensberg \latin{et~al.}(2021)Bensberg, Grimmel, Sobez, Steiner,
  Unsleber, and Reiher]{molassembler1.2.1}
Bensberg,~M.; Grimmel,~S.~A.; Sobez,~J.-G.; Steiner,~M.; Unsleber,~J.~P.;
  Reiher,~M. {qcscine/molassembler: Release 1.2.1}. 2021\relax
\mciteBstWouldAddEndPuncttrue
\mciteSetBstMidEndSepPunct{\mcitedefaultmidpunct}
{\mcitedefaultendpunct}{\mcitedefaultseppunct}\relax
\EndOfBibitem
\bibitem[Artymiuk \latin{et~al.}(1994)Artymiuk, Poirrette, Grindley, Rice, and
  Willett]{Artymiuk1994}
Artymiuk,~P.~J.; Poirrette,~A.~R.; Grindley,~H.~M.; Rice,~D.~W.; Willett,~P. {A
  graph-theoretic approach to the identification of three-dimensional patterns
  of amino acid side-chains in protein structures}. \emph{J. Mol. Biol.}
  \textbf{1994}, \emph{243}, 327--344\relax
\mciteBstWouldAddEndPuncttrue
\mciteSetBstMidEndSepPunct{\mcitedefaultmidpunct}
{\mcitedefaultendpunct}{\mcitedefaultseppunct}\relax
\EndOfBibitem
\bibitem[Spriggs \latin{et~al.}(2003)Spriggs, Artymiuk, and
  Willett]{Spriggs2003}
Spriggs,~R.~V.; Artymiuk,~P.~J.; Willett,~P. {Searching for patterns of amino
  acids in 3D protein structures}. \emph{{J. Chem. Inf. Comput. Sci.}}
  \textbf{2003}, \emph{43}, 412--421\relax
\mciteBstWouldAddEndPuncttrue
\mciteSetBstMidEndSepPunct{\mcitedefaultmidpunct}
{\mcitedefaultendpunct}{\mcitedefaultseppunct}\relax
\EndOfBibitem
\bibitem[Simm \latin{et~al.}(2020)Simm, T{\"{u}}rtscher, and Reiher]{Simm2020}
Simm,~G.~N.; T{\"{u}}rtscher,~P.~L.; Reiher,~M. {Systematic microsolvation
  approach with a cluster‐continuum scheme and conformational sampling}.
  \emph{J. Comput. Chem.} \textbf{2020}, \emph{41}, 1144--1155\relax
\mciteBstWouldAddEndPuncttrue
\mciteSetBstMidEndSepPunct{\mcitedefaultmidpunct}
{\mcitedefaultendpunct}{\mcitedefaultseppunct}\relax
\EndOfBibitem
\bibitem[Xue \latin{et~al.}(1998)Xue, {\"{O}}kvist, Hansson, and
  Young]{Xue1998}
Xue,~Y.; {\"{O}}kvist,~M.; Hansson,~{\"{O}}.; Young,~S. {Crystal structure of
  spinach plastocyanin at 1.7 \AA resolution}. \emph{Protein Sci.}
  \textbf{1998}, \emph{7}, 2099--2105\relax
\mciteBstWouldAddEndPuncttrue
\mciteSetBstMidEndSepPunct{\mcitedefaultmidpunct}
{\mcitedefaultendpunct}{\mcitedefaultseppunct}\relax
\EndOfBibitem
\bibitem[{IUPAC-IUB Commission on Biochemical
  Nomenclature}(1970)]{iupacAtomtyping}
{IUPAC-IUB Commission on Biochemical Nomenclature}, Abbreviations and Symbols
  for the Description of the Conformation of Polypeptide Chains: Tentative
  Rules (1969). \emph{Journal of Biological Chemistry} \textbf{1970},
  \emph{245}, 6489--6497\relax
\mciteBstWouldAddEndPuncttrue
\mciteSetBstMidEndSepPunct{\mcitedefaultmidpunct}
{\mcitedefaultendpunct}{\mcitedefaultseppunct}\relax
\EndOfBibitem
\bibitem[O'Boyle \latin{et~al.}(2011)O'Boyle, Banck, James, Morley,
  Vandermeersch, and Hutchison]{OpenBabel}
O'Boyle,~N.~M.; Banck,~M.; James,~C.~A.; Morley,~C.; Vandermeersch,~T.;
  Hutchison,~G.~R. {Open Babel: An open chemical toolbox}. \emph{J.
  Cheminform.} \textbf{2011}, \emph{3}, 33\relax
\mciteBstWouldAddEndPuncttrue
\mciteSetBstMidEndSepPunct{\mcitedefaultmidpunct}
{\mcitedefaultendpunct}{\mcitedefaultseppunct}\relax
\EndOfBibitem
\bibitem[Bensberg \latin{et~al.}(2022)Bensberg, Grimmel, Sobez, Steiner,
  Unsleber, and Reiher]{database1.0.0}
Bensberg,~M.; Grimmel,~S.~A.; Sobez,~J.-G.; Steiner,~M.; Unsleber,~J.~P.;
  Reiher,~M. {qcscine/database: Release 1.0.0}. 2022\relax
\mciteBstWouldAddEndPuncttrue
\mciteSetBstMidEndSepPunct{\mcitedefaultmidpunct}
{\mcitedefaultendpunct}{\mcitedefaultseppunct}\relax
\EndOfBibitem
\bibitem[Chang \latin{et~al.}(1997)Chang, J{\o}rgensen, Bardrum, and
  Led]{Chang1997}
Chang,~X.; J{\o}rgensen,~A. M.~M.; Bardrum,~P.; Led,~J.~J. {Solution structures
  of the R6 human insulin hexamer}. \emph{Biochemistry} \textbf{1997},
  \emph{36}, 9409--9422\relax
\mciteBstWouldAddEndPuncttrue
\mciteSetBstMidEndSepPunct{\mcitedefaultmidpunct}
{\mcitedefaultendpunct}{\mcitedefaultseppunct}\relax
\EndOfBibitem
\bibitem[Karlin and Zhu(1997)Karlin, and Zhu]{Zhu1997}
Karlin,~S.; Zhu,~Z.-Y. {Classification of mononuclear zinc metal sites in
  protein structures}. \emph{Proc. Natl. Acad. Sci.} \textbf{1997}, \emph{94},
  14231--14236\relax
\mciteBstWouldAddEndPuncttrue
\mciteSetBstMidEndSepPunct{\mcitedefaultmidpunct}
{\mcitedefaultendpunct}{\mcitedefaultseppunct}\relax
\EndOfBibitem
\bibitem[Rost and Sander(1993)Rost, and Sander]{Rost1993}
Rost,~B.; Sander,~C. {Prediction of protein secondary structure at better than
  70\% accuracy}. \emph{J. Mol. Biol.} \textbf{1993}, \emph{232},
  584--599\relax
\mciteBstWouldAddEndPuncttrue
\mciteSetBstMidEndSepPunct{\mcitedefaultmidpunct}
{\mcitedefaultendpunct}{\mcitedefaultseppunct}\relax
\EndOfBibitem
\bibitem[Cuff and Barton(1999)Cuff, and Barton]{Cuff1999}
Cuff,~J.~A.; Barton,~G.~J. {Evaluation and improvement of multiple sequence
  methods for protein secondary structure prediction}. \emph{Proteins Struct.
  Funct. Genet.} \textbf{1999}, \emph{34}, 508--519\relax
\mciteBstWouldAddEndPuncttrue
\mciteSetBstMidEndSepPunct{\mcitedefaultmidpunct}
{\mcitedefaultendpunct}{\mcitedefaultseppunct}\relax
\EndOfBibitem
\bibitem[Rost and Eyrich(2001)Rost, and Eyrich]{Rost2001}
Rost,~B.; Eyrich,~V.~A. {EVA: Large-scale analysis of secondary structure
  prediction}. \emph{Proteins Struct. Funct. Genet.} \textbf{2001}, \emph{45},
  192--199\relax
\mciteBstWouldAddEndPuncttrue
\mciteSetBstMidEndSepPunct{\mcitedefaultmidpunct}
{\mcitedefaultendpunct}{\mcitedefaultseppunct}\relax
\EndOfBibitem
\bibitem[Maier \latin{et~al.}(2015)Maier, Martinez, Kasavajhala, Wickstrom,
  Hauser, and Simmerling]{ff14sb}
Maier,~J.~A.; Martinez,~C.; Kasavajhala,~K.; Wickstrom,~L.; Hauser,~K.~E.;
  Simmerling,~C. {ff14SB: Improving the Accuracy of Protein Side Chain and
  Backbone Parameters from ff99SB}. \emph{Journal of Chemical Theory and
  Computation} \textbf{2015}, \emph{11}, 3696--3713\relax
\mciteBstWouldAddEndPuncttrue
\mciteSetBstMidEndSepPunct{\mcitedefaultmidpunct}
{\mcitedefaultendpunct}{\mcitedefaultseppunct}\relax
\EndOfBibitem
\bibitem[Thapa and Schlegel(2017)Thapa, and Schlegel]{Thapa2017a}
Thapa,~B.; Schlegel,~H.~B. {Improved pKa Prediction of Substituted Alcohols,
  Phenols, and Hydroperoxides in Aqueous Medium Using Density Functional Theory
  and a Cluster-Continuum Solvation Model}. \emph{J. Phys. Chem. A}
  \textbf{2017}, \emph{121}, 4698--4706\relax
\mciteBstWouldAddEndPuncttrue
\mciteSetBstMidEndSepPunct{\mcitedefaultmidpunct}
{\mcitedefaultendpunct}{\mcitedefaultseppunct}\relax
\EndOfBibitem
\bibitem[Thapa and Schlegel(2016)Thapa, and Schlegel]{Thapa2016}
Thapa,~B.; Schlegel,~H.~B. {Density Functional Theory Calculation of pKa's of
  Thiols in Aqueous Solution Using Explicit Water Molecules and the Polarizable
  Continuum Model}. \emph{J. Phys. Chem. A} \textbf{2016}, \emph{120},
  5726--5735\relax
\mciteBstWouldAddEndPuncttrue
\mciteSetBstMidEndSepPunct{\mcitedefaultmidpunct}
{\mcitedefaultendpunct}{\mcitedefaultseppunct}\relax
\EndOfBibitem
\bibitem[{Schmidt Am Busch} and Knapp(2004){Schmidt Am Busch}, and
  Knapp]{SchmidtAmBusch2004}
{Schmidt Am Busch},~M.; Knapp,~E.~W. {Accurate pKa determination for a
  heterogeneous group of organic molecules}. \emph{ChemPhysChem} \textbf{2004},
  \emph{5}, 1513--1522\relax
\mciteBstWouldAddEndPuncttrue
\mciteSetBstMidEndSepPunct{\mcitedefaultmidpunct}
{\mcitedefaultendpunct}{\mcitedefaultseppunct}\relax
\EndOfBibitem
\bibitem[Jurani{\'{c}}(2014)]{Juranic2014}
Jurani{\'{c}},~I. {Simple method for the estimation of pKa of amines}.
  \emph{Croat. Chem. Acta} \textbf{2014}, \emph{87}, 343--347\relax
\mciteBstWouldAddEndPuncttrue
\mciteSetBstMidEndSepPunct{\mcitedefaultmidpunct}
{\mcitedefaultendpunct}{\mcitedefaultseppunct}\relax
\EndOfBibitem
\bibitem[Beale \latin{et~al.}(2013)Beale, Chudzinski, Sarwar, and
  Taylor]{Beale2013}
Beale,~T.~M.; Chudzinski,~M.~G.; Sarwar,~M.~G.; Taylor,~M.~S. {Halogen bonding
  in solution: thermodynamics and applications}. \emph{Chem. Soc. Rev.}
  \textbf{2013}, \emph{42}, 1667--1680\relax
\mciteBstWouldAddEndPuncttrue
\mciteSetBstMidEndSepPunct{\mcitedefaultmidpunct}
{\mcitedefaultendpunct}{\mcitedefaultseppunct}\relax
\EndOfBibitem
\bibitem[Kromann \latin{et~al.}(2016)Kromann, Larsen, Moustafa, and
  Jensen]{Kromann2016}
Kromann,~J.~C.; Larsen,~F.; Moustafa,~H.; Jensen,~J.~H. {Prediction of pKa
  values using the PM6 semiempirical method}. \emph{PeerJ} \textbf{2016},
  \emph{2016}, e2335\relax
\mciteBstWouldAddEndPuncttrue
\mciteSetBstMidEndSepPunct{\mcitedefaultmidpunct}
{\mcitedefaultendpunct}{\mcitedefaultseppunct}\relax
\EndOfBibitem
\bibitem[Jensen \latin{et~al.}(2017)Jensen, Swain, and Olsen]{Jensen2017}
Jensen,~J.~H.; Swain,~C.~J.; Olsen,~L. {Prediction of p K a Values for Druglike
  Molecules Using Semiempirical Quantum Chemical Methods}. \emph{J. Phys. Chem.
  A} \textbf{2017}, \emph{121}, 699--707\relax
\mciteBstWouldAddEndPuncttrue
\mciteSetBstMidEndSepPunct{\mcitedefaultmidpunct}
{\mcitedefaultendpunct}{\mcitedefaultseppunct}\relax
\EndOfBibitem
\bibitem[Hunt \latin{et~al.}(2020)Hunt, Hosseini-Gerami, Chrien, Plante,
  Ponting, and Segall]{Hunt2020}
Hunt,~P.; Hosseini-Gerami,~L.; Chrien,~T.; Plante,~J.; Ponting,~D.~J.;
  Segall,~M. {Predicting p K a Using a Combination of Semi-Empirical Quantum
  Mechanics and Radial Basis Function Methods}. \emph{J. Chem. Inf. Model.}
  \textbf{2020}, \emph{60}, 2989--2997\relax
\mciteBstWouldAddEndPuncttrue
\mciteSetBstMidEndSepPunct{\mcitedefaultmidpunct}
{\mcitedefaultendpunct}{\mcitedefaultseppunct}\relax
\EndOfBibitem
\bibitem[Fitch \latin{et~al.}(2002)Fitch, Karp, Lee, Stites, Lattman, and
  Garc\'{\i}a-Moreno]{Fitch2002}
Fitch,~C.~A.; Karp,~D.~A.; Lee,~K.~K.; Stites,~W.~E.; Lattman,~E.~E.;
  Garc\'{\i}a-Moreno,~E.~B. {Experimental pKa Values of Buried Residues:
  Analysis with Continuum Methods and Role of Water Penetration}.
  \emph{Biophys. J.} \textbf{2002}, \emph{82}, 3289--3304\relax
\mciteBstWouldAddEndPuncttrue
\mciteSetBstMidEndSepPunct{\mcitedefaultmidpunct}
{\mcitedefaultendpunct}{\mcitedefaultseppunct}\relax
\EndOfBibitem
\bibitem[Thurlkill \latin{et~al.}(2006)Thurlkill, Grimsley, Scholtz, and
  Pace]{Thurlkill2005}
Thurlkill,~R.~L.; Grimsley,~G.~R.; Scholtz,~J.~M.; Pace,~C.~N. {pK values of
  the ionizable groups of proteins}. \emph{Protein Science} \textbf{2006},
  \emph{15}, 1214--1218\relax
\mciteBstWouldAddEndPuncttrue
\mciteSetBstMidEndSepPunct{\mcitedefaultmidpunct}
{\mcitedefaultendpunct}{\mcitedefaultseppunct}\relax
\EndOfBibitem
\bibitem[Martin \latin{et~al.}(1999)Martin, Richard, Salem, Hartley, and
  Mauguen]{Martin1996}
Martin,~C.; Richard,~V.; Salem,~M.; Hartley,~R.; Mauguen,~Y. {Refinement and
  structural analysis of barnase at 1.5{\AA} resolution}. \emph{Acta
  Crystallographica Section D} \textbf{1999}, \emph{55}, 386--398\relax
\mciteBstWouldAddEndPuncttrue
\mciteSetBstMidEndSepPunct{\mcitedefaultmidpunct}
{\mcitedefaultendpunct}{\mcitedefaultseppunct}\relax
\EndOfBibitem
\bibitem[Oliveberg \latin{et~al.}(1995)Oliveberg, Arcus, and
  Fersht]{Oliveberg1995}
Oliveberg,~M.; Arcus,~V.~L.; Fersht,~A.~R. {pKA Values of Carboxyl Groups in
  the Native and Denatured States of Barnase: The pKA Values of the Denatured
  State Are on Average 0.4 Units Lower Than Those of Model Compounds}.
  \emph{Biochemistry} \textbf{1995}, \emph{34}, 9424--9433\relax
\mciteBstWouldAddEndPuncttrue
\mciteSetBstMidEndSepPunct{\mcitedefaultmidpunct}
{\mcitedefaultendpunct}{\mcitedefaultseppunct}\relax
\EndOfBibitem
\bibitem[Loewenthal \latin{et~al.}(1991)Loewenthal, Sancho, and
  Fersht]{Loewenthal1991}
Loewenthal,~R.; Sancho,~J.; Fersht,~A.~R. {Fluorescence spectrum of barnase:
  contributions of three tryptophan residues and a histidine-related pH
  dependence}. \emph{Biochemistry} \textbf{1991}, \emph{30}, 6775--6779\relax
\mciteBstWouldAddEndPuncttrue
\mciteSetBstMidEndSepPunct{\mcitedefaultmidpunct}
{\mcitedefaultendpunct}{\mcitedefaultseppunct}\relax
\EndOfBibitem
\bibitem[Pahari \latin{et~al.}(2019)Pahari, Sun, and Alexov]{pkad}
Pahari,~S.; Sun,~L.; Alexov,~E. {PKAD: a database of experimentally measured
  pKa values of ionizable groups in proteins.} 2019\relax
\mciteBstWouldAddEndPuncttrue
\mciteSetBstMidEndSepPunct{\mcitedefaultmidpunct}
{\mcitedefaultendpunct}{\mcitedefaultseppunct}\relax
\EndOfBibitem
\end{mcitethebibliography}
\providecommand{\latin}[1]{#1}
\makeatletter
\providecommand{\doi}
  {\begingroup\let\do\@makeother\dospecials
  \catcode`\{=1 \catcode`\}=2 \doi@aux}
\providecommand{\doi@aux}[1]{\endgroup\texttt{#1}}
\makeatother
\providecommand*\mcitethebibliography{\thebibliography}
\csname @ifundefined\endcsname{endmcitethebibliography}
  {\let\endmcitethebibliography\endthebibliography}{}

%

\end{document}